\documentclass[preprint2]{aastex}
\usepackage{amsmath}
\usepackage{booktabs}
\usepackage{longtable}
\usepackage{xspace}
\usepackage{hyperref}
\usepackage[T1]{fontenc}
\usepackage{lipsum}
\usepackage{comment}

\usepackage{natbib}
\usepackage{graphicx}
\newcommand{\kms}{\ensuremath{\rm km\,s^{-1}}}

\newcommand{\teff}{\ensuremath{T_{\rm eff}}}
\newcommand{\logg}{\ensuremath{\log{g}}}
\newcommand{\vsini}{\ensuremath{v \sin{i}}}
\newcommand{\feh}{[Fe/H]}

\newcommand{\rsun}{\ensuremath{R_\sun}}
\newcommand{\msun}{\ensuremath{M_\sun}}

\newcommand{\rstar}{\ensuremath{R_\star}}
\newcommand{\mstar}{\ensuremath{M_\star}}

\newcommand{\rhostar}{\ensuremath{\rho_\star}}

\newcommand{\rpl}{\ensuremath{R_{\rm p}}}

\newcommand{\rearth}{\ensuremath{R_\earth}}

\newcommand{\msini}{\ensuremath{m \sin{i}}}

\newcommand{\Kepler}{\textit{Kepler}}
\newcommand{\npl}{892\xspace}
\newcommand{\nstars}{349\xspace}
\newcommand{\rthresh}{\ensuremath{1.5~\rearth}}
\newcommand{\Pthresh}{\ensuremath{30~\mathrm{days}}}
\newcommand{\Ntp}{\ensuremath{N_\mathrm{tp}}}

\newcommand{\nsingles}{376\xspace}
\newcommand{\nmultip}{426\xspace}
\newcommand{\ndoublep}{446\xspace}
\newcommand{\ntriplep}{228\xspace}
\newcommand{\nfourplup}{218\xspace}

\newcommand{\nmultis}{166\xspace}
\newcommand{\ndoubles}{223\xspace}
\newcommand{\ntriples}{76\xspace}
\newcommand{\nfourplus}{50\xspace}

\submitjournal{The Astronomical Journal}
\accepted{October 01, 2018}

\begin{document}
\title{The California-Kepler Survey. VI:  \Kepler\ Multis and Singles Have Similar Planet and Stellar Properties Indicating a Common Origin\footnote{Based on observations obtained at the W.\,M.\,Keck Observatory, which is operated jointly by the University of California and the California Institute of Technology.  Keck time was granted for this project by the University of California, and California Institute of Technology, the University of Hawaii, and NASA.}}
\author[0000-0002-3725-3058]{Lauren M. Weiss}
\affiliation{Institute for Astronomy, University of Hawaii at Manoa, Honolulu, HI, USA}
\affiliation{Parrent Fellow}
\affiliation{Universit{\'e} de Montr{\'e}al, Montr{\'e}al, QC, Canada}
\affiliation{Trottier Fellow}

\author[0000-0002-0531-1073]{Howard T. Isaacson}
\affiliation{University of California at Berkeley, Berkeley, CA, USA}

\author[0000-0002-2909-0113]{Geoffrey W. Marcy}
\affiliation{University of California at Berkeley, Berkeley, CA, USA}

\author[0000-0001-8638-0320]{Andrew W. Howard} 
\affiliation{California Institute of Technology, Pasadena, CA, USA}

\author[0000-0003-0967-2893]{Erik A. Petigura}
\affiliation{California Institute of Technology, Pasadena, CA, USA}
\affiliation{NASA Sagan Fellow}

\author[0000-0003-3504-5316]{Benjamin J. Fulton}
\affiliation{NASA Exoplanet Science Institute, Pasadena, CA, USA}
\affiliation{California Institute of Technology, Pasadena, CA, USA}
\affiliation{Texaco Fellow}

\author[0000-0002-4265-047X]{Joshua N. Winn}
\affiliation{Princeton University, Princeton, NJ, USA}

\author[0000-0001-8058-7443]{Lea Hirsch}
\affiliation{University of California at Berkeley, Berkeley, CA, USA}

\author[0000-0002-5658-0601]{Evan Sinukoff}
\affiliation{Institute for Astronomy, University of Hawaii at Manoa, Honolulu, HI, USA}
\affiliation{NSERC Graduate Research Fellow}

\author[0000-0002-5904-1865]{Jason F. Rowe}
\affiliation{Bishops University, Sherbrooke, QC, Canada}

\collaboration{The California Kepler Survey}

\begin{abstract}
The California-Kepler Survey (CKS) catalog contains precise stellar and planetary properties for the \Kepler\ planet candidates, including systems with multiple detected transiting planets (``multis'') and systems with just one detected transiting planet (``singles,'' although additional planets could exist).  We compared the stellar and planetary properties of the multis and singles in a homogenous subset of the full CKS-Gaia catalog.  We found that sub-Neptune sized singles and multis do not differ in their stellar properties or planet radii.  In particular: 
(1.) The distributions of stellar properties \mstar, [Fe/H], and \vsini\ for the \Kepler\ sub Neptune-sized singles and multis are statistically indistinguishable.
(2.) The radius distributions of the sub-Neptune sized singles and multis with $P > 3$ days are indistinguishable, and both have a valley at $\sim1.8~\rearth$.  However, 
there are significantly more detected short-period ($P < 3$ days), sub-Neptune sized singles than multis.
The similarity of the host star properties, planet radii, and radius valley for singles and multis suggests a common origin.  The similar radius valley, which is likely sculpted by photo-evaporation from the host star within the first 100 Myr, suggests that planets in both singles and multis spend much of the first 100 Myr near their present, close-in locations.  One explanation that is consistent with the similar fundamental properties of singles and multis is that many of the singles are members of multi-planet systems that underwent planet-planet scattering.
\end{abstract}


\section{Introduction}
Comparisons between planetary systems with multiple planets and those with just one known planet have long been used to probe planet formation.  A decade after the discovery of the first multi-planet system around a main sequence star \citep{Butler1999}, \citet{Wright2009} conducted a statistical study of 28 multi-planet systems, all of which were discovered and characterized with radial velocities.  They compared the multi-planet systems to systems with only one known planet and found that multi-planet systems were spaced uniformly in log-period (unlike the single-planet systems) and typically had lower eccentricities and \msini\ values than the single-planet systems.

More recently, the \Kepler\ Mission \citep{Borucki2010} has detected hundreds of multi-planet systems \citep{Latham2011,Lissauer2011_multis,Fabrycky2014,Lissauer2014,Rowe2014}.  In the \Kepler\ multi-planet systems, multiple planet candidates transit the star, resulting in measured orbital periods, planet-to-star radius ratios, and transit durations for each planet. The vast majority of the \Kepler\ planet candidates in multis are bona-fide planets, based on statistical arguments \citep{Lissauer2012,Lissauer2014}.  The \Kepler\ multi-planet systems differ from the previously studied RV multi-planet systems in that \Kepler\ was sensitive to smaller (lower-mass) planets.  The majority of the \Kepler\ single-planet and multi-planet systems have sub-Neptune sized planets rather than giant planets \citep{Latham2011}.  Also, \Kepler\ only detected transiting planets.  

In systems with multiple transiting planets, the planets are very likely nearly coplanar by virtue of the fact that they all transit \citep{Lissauer2011_multis}.  However, not all multi-planet systems must be nearly coplanar.  A sufficiently non-coplanar system might result in only one transiting planet detected by \Kepler, although multiple planets might exist.  The systems with just one detected transiting planet (``singles'') might belong to the tail of a single underlying distribution that describes systems with multiple detected transiting planets (``multis'').  On the other hand, a high fraction of the singles might belong to a population with different formation conditions or a different dynamical history.

We would like to understand whether the \Kepler\ singles and multis differ in their orbital and physical parameters.  Some orbital parameters of interest include multiplicity, orbital periods, eccentricities, and inclinations.  Physical parameters of interest include host star mass, metallicity, and rotation velocity, as well as planet radius and mass.  If the singles differ from the multis in their underlying distributions of orbital and/or physical parameters, such a distinction likely points to a divergence in the planet formation and/or evolution of the \Kepler\ singles versus multis.

Past research has considered the hypothesis that a large fraction of the \Kepler\ singles belong to a distinct population from the multi-planet systems.  Some examples of a distinct population are a dynamically hot population (high mutual inclinations and eccentricities for the singles) or a population with wider spacing in the orbital period ratios for the singles than is typical for the multis.  \citet{Lissauer2011_multis} found that the typical mutual inclinations in the \Kepler\ multis were $< 10^\circ$ and noted that these small mutual inclinations seemed inconsistent with the large number of observed singles.  \citet{Hansen2013} explored the multiplicity vectors and period distributions of the \Kepler\ singles and multis through a model of \textit{in situ} planet formation.  They found that the number of \Kepler\ singles is too high to result from an in situ formation scenario (although the authors required each system to have at least three initially coplanar planets).  In another study that required a minimum number of planets per system, \citet{Ballard2016} found that there is an excess of singles among the \Kepler\ M-dwarfs.  \citet{Xie2016} used stellar spectra from LAMOST and the transit durations from \Kepler\ lightcurves to estimate of the mean eccentricities and inclinations for singles and multis.  They found that the mean eccentricity of the singles was $\sim0.3$, whereas the multis were on nearly circular orbits ($e = 0.04\pm0.04$).  In a sample of stars with asteroseismically determined properties, \citet{vanEylen2018_ecc} also found higher eccentricities for the singles than the multis.

However, other studies have found no need for a large fraction of the singles to have distinct underlying architectures.   \citet{Ford2011} found that the prevalence of TTVs in singles was consistent with the multis, suggesting that many singles belong to compact, multi-planet systems.
\citet{Tremaine2012} explored a variety of possible orbital geometries and found that no separate population was needed to explain the apparent excess of \Kepler\ singles, if high mutual inclinations were allowed in a small fraction of the multis.  \citet{Fang2012} modeled the transit duration ratios as well as the transiting planet multiplicity.  They found that an underlying distribution in which most multi-planet systems have mutual inclination distributions of $<3^\circ$, and 75\% of systems have 1-2 planets with $P < 200$ days (like the solar system) describes the observed planet multiplicities and transit duration ratios.  \citet{Gaidos2016} found that with improved stellar parameters and an exponentially-distributed number of planets per star, the large number of M dwarf singles compared to multis announced in \citet{Ballard2016} could be reconciled.  \citet{Zhu2018} used spectra from LAMOST to measure the properties of \Kepler\ planet candidate host stars, giving special attention to the differences between multis and singles that did and did not exhibit transit timing variations.  They found that the stellar properties of the singles and multis did not differ substantially.  \citet{MunozRomero2018} compared the metallicities determined by the California-\textit{Kepler} Survey (described below) for singles and multis and found no significant differences in the stellar metallicities.

We push the comparison of the fundamental properties of the \Kepler\ singles versus multis into new regions of parameter space by leveraging the precise stellar and planetary parameters of The California-\Kepler\ Survey (CKS) combined with Gaia DR2.  CKS obtained high-resolution (R=60,000) spectra for 1305 \Kepler\ systems with transiting planets \citep{Petigura2017}.  The improved stellar and planetary parameters \citep[][CKS II]{Johnson2017} enable a more accurate and precise characterization of the \Kepler\ systems than was previously available, yielding 2025 transiting planet candidates with precise radii and host star properties.  Fulton \& Petigura 2018 (CKS VII) revised the stellar properties and planet radii based on parallaxes from the Gaia DR2 catalog \citep{GaiaDR2}.  CKS and Gaia have dramatically improved the characterization of the \Kepler\ stellar radii, metallicities, masses, and rotations, as well as the planet radii and equilibrium temperatures, compared to what was available before the CKS project \citep[e.g.,][]{Brown2011}.

In this paper (CKS VI), we use the refined stellar and planetary properties presented in CKS VII to compare a large, homogeneous, high-purity sample of \Kepler\ singles and multis.  Where applicable, we also examine how the stellar and planetary properties of the multis differ for system with 2, 3, and 4 or more transiting planets.

In section \ref{sec:sample}, we discuss the cuts to the CKS catalog needed to generate homogenous samples for comparison.  In section \ref{sec:stars}, we compare the distributions of the stellar properties for the singles vs. the multis.  In section \ref{sec:planets}, we compare the distributions of the planet radii and orbital periods for the singles vs. the multis.  We conclude in section \ref{sec:conclusion}.


\section{The Sample}
\label{sec:sample}
By construction, CKS was not a homogenous survey \citep{Petigura2017}.  The largest component of CKS is \Kepler\ planet hosts with $Kp < 14.2$.  However, CKS was expanded to include fainter \Kepler\ stars that addressed special interests, including ultra-short period planets, planets in the habitable zone, and multi-planet systems.

Since this paper addresses multi-planet systems, we would like to include the full breadth of multi-planet systems wherever possible.  However, the population of CKS singles is heterogenous: most orbit stars with $Kp < 14.2$, and any singles orbiting fainter stars are ultra-short period planets or habitable-zone planets.  Thus, the sample of singles is only homogenous for $Kp < 14.2$.  

The different magnitude limits of homogenous sub-samples of the singles and multis are problematic because fundamental stellar properties, in particular stellar mass and radius, are correlated with stellar magnitude.  Therefore, to fairly compare the singles and multis, we must down-select the multis to those with $Kp < 14.2$.  However, in comparing multis to each other, we can use the full sample of CKS-Gaia.

In addition to ensuring homogenous samples for comparison, we make several cuts to ensure high-precision stellar and planetary parameters.

\subsection{Selecting High-Purity Planet Samples}
The CKS planet candidates and the successive cuts we made are summarized in Table \ref{tab:cuts}.  The initial CKS dataset consists of 1944 signals that were at one time flagged as transiting planet candidates, orbiting 1222 stars that have Gaia properties reported in DR2.  From these, we discarded the signals that are now known to be false positives (as determined on either the NASA Exoplanet Archive or in CKS I), removing 156 non-planetary signals around 104 stars.  We then discarded stars that are diluted by at least 5\% by a second star in the \Kepler\ aperture \citep[as determined in the stellar companion catalog of][]{Furlan2017}, removing 88 planet candidates around 58 stars.  We discarded planets for which \citet{Mullally2015} measured $b > 0.9$, for which the high impact parameters adversely affected our ability to determine accurate planet radii, removing 137 planet candidates from around 70 stars.  We removed planet candidates for which the measured signal-to-noise ratio (SNR) is less than 10 since these planets have poorly determined radii and impact parameters, removing 48 planet candidates.  We also removed planet candidates with $\rpl > 22.4~\rearth$, which are likely eclipsing binaries rather than planets.  Of the four planet candidates with $\rpl > 22.4~\rearth$, all four were singles, and three orbited giant stars with $\logg < 3.9$.  Systems that were originally multis but had been purified to the extent that only one planet remained were excluded.  

After these cuts, our sample included \npl high-purity planet candidates in multi-planet systems around \nstars stars\footnote{This number differs slightly from \citep{Weiss2018} because a few stars from that study did not have parallaxes in Gaia DR2.}.  The number of planets and stars in the various subsamples are summarized in Table \ref{tab:sample}.  In this sample of multis, we compare the systems with 2 transiting planets (\ndoublep planets around \ndoubles stars), 3 transiting planets (\ntriplep planets around \ntriples stars), and 4+ transiting planets (\nfourplup planets around \nfourplus stars).  Figure \ref{fig:npl} shows the number of stars with various multiplicities (blue histogram).  We present the catalog of planets in the high-purity sample of multis in Table \ref{tab:cks2-short}.

\begin{deluxetable}{cccc}
\tablecaption{Successive Cuts\label{tab:cuts}}
\tablehead{\colhead{\Ntp\tablenotemark{a}} & \colhead{$N_\star$} & \colhead{$N_\mathrm{tp,multi}$} & \colhead{Cut}} 
\startdata
1944 &  1222 &  1176 &  \\
1788 &  1118 &  1092 &  No FPs \\
1700 &  1060 &  1042 &  dilution $< 5\%$ \\
1563 &  990 &  940 &  $b < $ 0.9 \\
1563 &  990 &  940 & $\rpl/\rstar < $ 0.5 \\
1495 &  952 &  908 & SNR $ > $ 10.0 \\
1491 &  948 & 892\tablenotemark{b} &  $\rpl < $ 22.4 \\
997 &  700 &  492 & $ Kp < $ 14.2 \\
843 &  578 &  434\tablenotemark{c} & SNR $1.5~\rearth$, 30 days $>$ 10.0\\
\enddata
\tablenotetext{a}{Number of transiting planets}
\tablenotetext{b}{These are the ``CKS Multis'' sample}
\tablenotetext{c}{These are the $\mathcal{B}_m$ sample}
\end{deluxetable}

\subsection{Selecting Singles and Multis for Comparison}
\begin{figure}
\begin{center}
\includegraphics[width=\columnwidth]{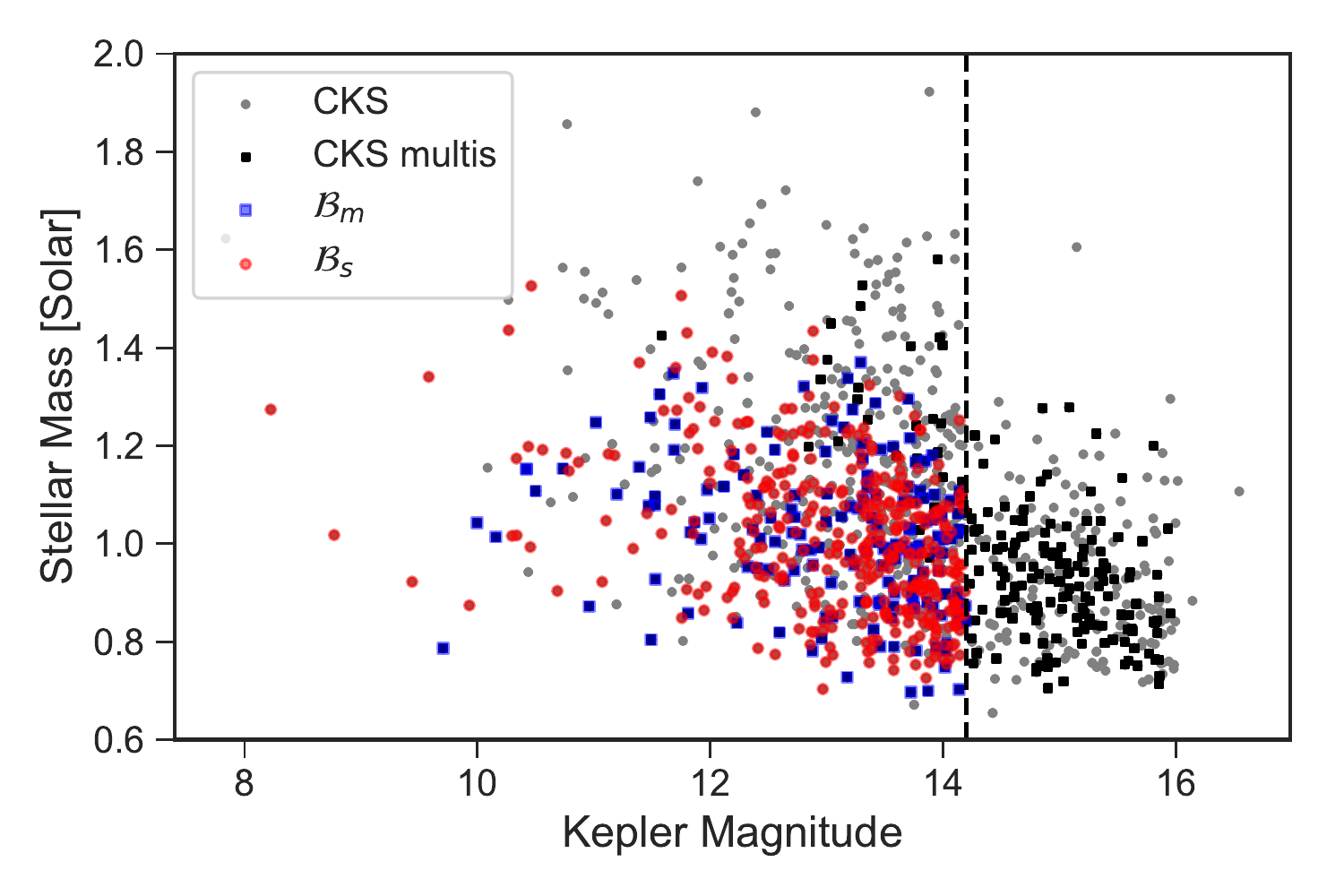}
\caption{Stellar mass vs. Kepler magnitude for the purified CKS-Gaia singles (gray circles) and multis (black squares), and the homogenized, magnitude-limited singles (red) and multis (blue).  There is a correlation between stellar mass and magnitude, motivating a homogenous, magnitude-limited sample of singles and multis to ensure a fair comparison of their host star properties.}
\label{fig:malmquist}
\end{center}
\end{figure}

As discussed above, a homogenous comparison of singles and multis can only be performed for $Kp < 14.2.$  In the CKS sample, we observe a significant correlation between \Kepler\ magnitude and stellar mass (Pearson $r = -3$, $p < 10^{-5}$, see Figure \ref{fig:malmquist}).  This is in part because host star apparent magnitude correlates with luminosity (Malmquist 1922), which correlates with both stellar mass and radius.  In addition, the 150,000 stars selected for monitoring in the \Kepler\ mission were chosen based on both their magnitudes and their colors (as a proxy for spectral type), and so the selection criteria might have contributed to the correlation.  Hence, a common apparent magnitude cut for the singles and multis ensures that any observed difference in the host star properties is astrophysical rather than the result of selection biases.  From the multis in the purified sample, we  down-selected to those orbiting stars with $Kp < 14.2$, resulting in $\nmultip$ planets orbiting $\nmultis$ stars.

\begin{figure}
\begin{center}
\includegraphics[width=\columnwidth]{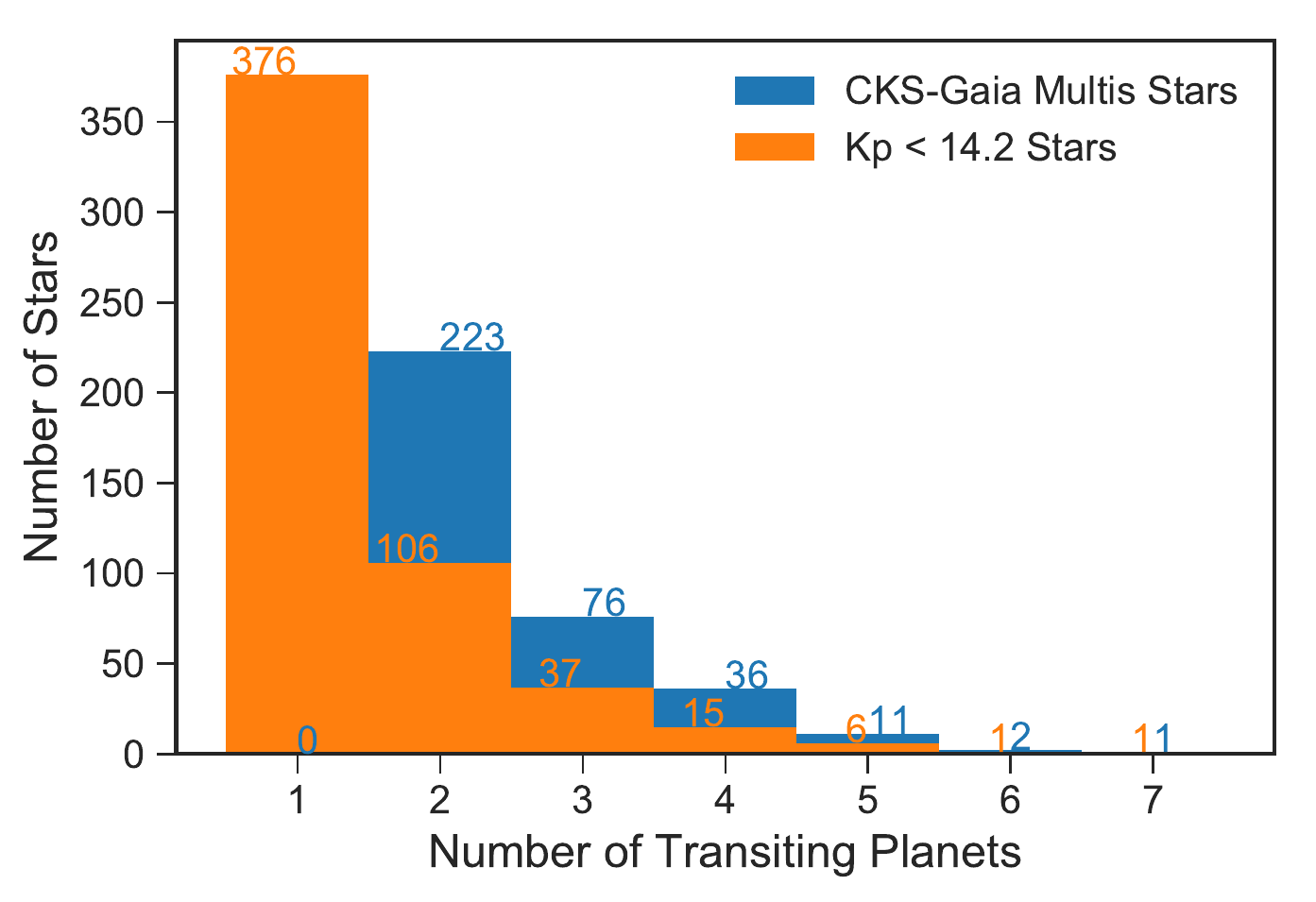}
\caption{Histograms of the number of stars with various transiting planet multiplicities in the full CKS-Gaia multis sample (blue) and in the magnitude-limited $\mathcal{B}_s + \mathcal{B}_m$ samples (orange).}
\label{fig:npl}
\end{center}
\end{figure}

Among both the singles and multis, we did not want to include systems for which it would have been unlikely to detect additional planets, and so we made a final cut based on the detectability of a hypothetical planet around each star in our sample.  To determine the ease of detecting planets around a given star, we calculated the model signal-to-noise ratio (SNR) for a \rthresh\ planet orbiting at \Pthresh\ around each of the single and multi host stars via the following equations:
\begin{equation}
\mathrm{SNR} = \frac{(\rpl/\rstar)^2 \sqrt{3.5 \mathrm{yr}/P}}{\mathrm{CDPP_{6h}} \sqrt{6 \mathrm{hr} / T}}
\label{eqn:detectability}
\end{equation}
\begin{equation}
T = 13\mathrm{hr}~(P/1 \mathrm{yr})^{1/3} (\rhostar/\rho_\odot)^{-1/3}
\end{equation}

where $\rpl$ is the planet radius, $\rstar$ is the stellar radius, $\rhostar/\rho_\odot$ is the stellar density in units of solar density, and $\mathrm{CDPP_{6h}}$ is the combined differential photometric precision in the \Kepler\ light curve over 6 hours.  We remove all singles \textit{and multis} stars for which the model SNR $< 10$, ensuring that a planet of \rthresh\ at \Pthresh\ would have been detectable around all the stars in our sample.  Our final sample of $Kp < 14.2$ singles, hereafter $\mathcal{B}_s$, contains \nsingles singles.  Our final sample of $Kp < 14.2$ multis, hereafter $\mathcal{B}_m$, contains \nmultis\ stars hosting \nmultip\ planets.  The $\mathcal{B}_m$ and $\mathcal{B}_s$ samples together comprise the orange histogram in Figure \ref{fig:npl}.  The planet and stellar properties of the $\mathcal{B}_m$ and $\mathcal{B}_s$ samples are listed in Table \ref{tab:multi-v-single}.

\begin{deluxetable*}{cccccccccccc}



\tablewidth{0pt}

\tablecaption{High Purity CKS-Gaia Multis \label{tab:cks2-short}}


\tablehead{\colhead{KOI} & \colhead{Kepmag} & \colhead{CDPP$_\mathrm{6h}$} & \colhead{\teff} & \colhead{\logg} & \colhead{[Fe/H]} & \colhead{\vsini} & \colhead{$\mstar$} & \colhead{$\rstar$} & \colhead{Period} & \colhead{$\rpl$} & \colhead{$N_\mathrm{tp}$}\\ 
\colhead{} & \colhead{} & \colhead{(ppm)} & \colhead{(K)} & \colhead{} & \colhead{} & \colhead{(\kms)} & \colhead{(\msun)} & \colhead{(\rsun)} & \colhead{(days)} & \colhead{(\rearth)} & \colhead{\tablenotemark{a}} } 
\startdata
K00041.01 & 11.2 & 23.33 & 5873.54 & 4.11 & 0.09 & 2.7 & 1.1 & 1.54 & 12.82 & 2.36 & 3 \\
K00041.02 & 11.2 & 23.33 & 5873.54 & 4.11 & 0.09 & 2.7 & 1.1 & 1.54 & 6.89 & 1.35 & 3 \\
K00041.03 & 11.2 & 23.33 & 5873.54 & 4.11 & 0.09 & 2.7 & 1.1 & 1.54 & 35.33 & 1.54 & 3 \\
K00046.01 & 13.77 & 54.61 & 5686.15 & 4.06 & 0.38 & 2.5 & 1.24 & 1.72 & 3.49 & 6.19 & 2 \\
K00046.02 & 13.77 & 54.61 & 5686.15 & 4.06 & 0.38 & 2.5 & 1.24 & 1.72 & 6.03 & 1.29 & 2 \\
K00070.01 & 12.5 & 39.42 & 5483.75 & 4.51 & 0.08 & 0 & 0.95 & 0.89 & 10.85 & 2.93 & 5 \\
K00070.02 & 12.5 & 39.42 & 5483.75 & 4.51 & 0.08 & 0 & 0.95 & 0.89 & 3.7 & 2.04 & 5 \\
K00070.03 & 12.5 & 39.42 & 5483.75 & 4.51 & 0.08 & 0 & 0.95 & 0.89 & 77.61 & 2.53 & 5 \\
K00070.04 & 12.5 & 39.42 & 5483.75 & 4.51 & 0.08 & 0 & 0.95 & 0.89 & 6.1 & 0.8 & 5 \\
K00070.05 & 12.5 & 39.42 & 5483.75 & 4.51 & 0.08 & 0 & 0.95 & 0.89 & 19.58 & 0.96 & 5 \\
\enddata
\tablenotetext{a}{Number of transiting planets surviving our cuts.}

\tablecomments{This table is downloadable in full online.  A portion has been reproduced here for form and content.}


\end{deluxetable*}

\begin{deluxetable*}{ccccccccccccc}



\tablewidth{0pt}

\tablecaption{Bright Multis and Singles \label{tab:multi-v-single}}


\tablehead{\colhead{KOI} & \colhead{Kepmag} & \colhead{CDPP$_\mathrm{6h}$} & \colhead{\teff} & \colhead{\logg} & \colhead{[Fe/H]} & \colhead{\vsini} & \colhead{$\mstar$} & \colhead{$\rstar$} & \colhead{Period} & \colhead{$\rpl$} & \colhead{Model SNR} & \colhead{$N_\mathrm{tp}$}\\ 
\colhead{} & \colhead{} & \colhead{(ppm)} & \colhead{(K)} & \colhead{} & \colhead{} & \colhead{(\kms)} & \colhead{(\msun)} & \colhead{(\rsun)} & \colhead{(days)} & \colhead{(\rearth)} & \colhead{\tablenotemark{a}} & \colhead{\tablenotemark{b}} } 

\startdata
K00001.01 & 11.34 & 17.72 & 5820.3 & 4.39 & -0.01 & 1.3 & 0.99 & 1.05 & 2.47 & 14.24 & 61.72 & 1 \\
K00002.01 & 10.46 & 21.36 & 6448.66 & 4.02 & 0.18 & 5.2 & 1.53 & 2 & 2.2 & 16.44 & 25.56 & 1 \\
K00007.01 & 12.21 & 30.05 & 5844.82 & 4.13 & 0.17 & 2.8 & 1.16 & 1.54 & 3.21 & 4.16 & 20.67 & 1 \\
K00017.01 & 13.3 & 42.06 & 5664.08 & 4.26 & 0.34 & 2.6 & 1.1 & 1.28 & 3.23 & 13.35 & 16.02 & 1 \\
K00018.01 & 13.37 & 46.12 & 6326.65 & 4.08 & 0 & 4.4 & 1.32 & 1.74 & 3.55 & 15.25 & 12.41 & 1 \\
K00020.01 & 13.44 & 51.55 & 5945.39 & 4.1 & 0.02 & 1.7 & 1.1 & 1.56 & 4.44 & 20.12 & 10.54 & 1 \\
K00022.01 & 13.44 & 50.59 & 5880.28 & 4.26 & 0.18 & 1.9 & 1.13 & 1.29 & 7.89 & 13.28 & 14.31 & 1 \\
K00041.01 & 11.2 & 23.33 & 5873.54 & 4.11 & 0.09 & 2.7 & 1.1 & 1.54 & 12.82 & 2.36 & 26.16 & 3 \\
K00041.02 & 11.2 & 23.33 & 5873.54 & 4.11 & 0.09 & 2.7 & 1.1 & 1.54 & 6.89 & 1.35 & 26.16 & 3 \\
K00041.03 & 11.2 & 23.33 & 5873.54 & 4.11 & 0.09 & 2.7 & 1.1 & 1.54 & 35.33 & 1.54 & 26.16 & 3 \\
\enddata
\tablenotetext{a}{Model SNR for a \rthresh\ planet at \Pthresh.}
\tablenotetext{b}{Number of transiting planets surviving our cuts.}

\tablecomments{This table is downloadable in full online.  A portion has been reproduced here for form and content.}




\end{deluxetable*}

\begin{figure}
\begin{center}
\includegraphics[width=\columnwidth]{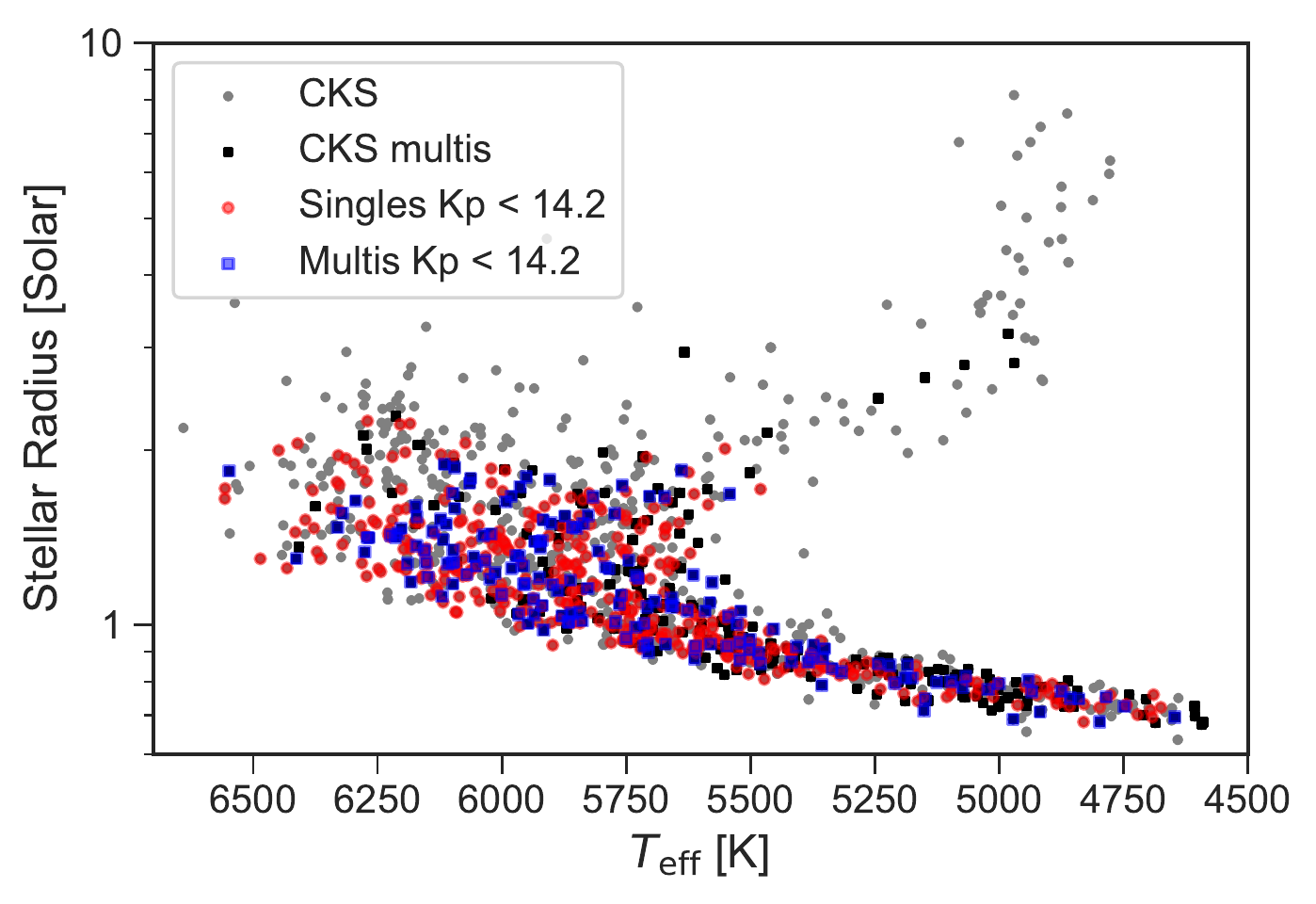}
\caption{Stellar radius vs. effective temperature for the purified CKS-Gaia singles (gray circles) and multis (black squares), and the homogenized, magnitude-limited singles (red) and multis (blue).}
\label{fig:hr}
\end{center}
\end{figure}

We tabulate the number of stars and planets in the initial CKS-Gaia sample, the cleaned sample of multis, the subsets with 2, 3, or 4+ transiting planets (labeled $\Ntp=2$, $\Ntp=3$, and $\Ntp \ge 4$), and the $Kp < 14.2$ singles and multis (labeled $\mathcal{B}_s$ and $\mathcal{B}_m$) in Table \ref{tab:sample}.  In addition, we include the subsets of $\mathcal{B}_s$ and $\mathcal{B}_m$ in which the planets are smaller than 4~\rearth\ ($\mathcal{B}_{s,4}$ and $\mathcal{B}_{m,4}$).  The host star effective temperatures and radii of the CKS stars, the high-purity multis, $\mathcal{B}_s$, and $\mathcal{B}_m$ are summarized in Figure \ref{fig:hr}.

\begin{deluxetable}{lcrrl}
\tablecaption{The Samples\label{tab:sample}}
\tablehead{\colhead{Name} & \colhead{Description} & \colhead{$\Ntp$\tablenotemark{a}} & \colhead{$N_\star$}}
\startdata
CKS Multis & High-Purity CKS Multis& \npl & \nstars \\
\hline
$N_\mathrm{tp}=2$ & 2 transiting planets & \ndoublep & \ndoubles \\
$N_\mathrm{tp}=3$ & 3 transiting planets & \ntriplep & \ntriples  \\
$N_\mathrm{tp}\ge4$ & 4+ transiting planets & \nfourplup & \nfourplus \\
\hline
$\mathcal{B}_{s}$ & Bright ($Kp < 14.2\tablenotemark{b}$) Singles & \nsingles & \nsingles \\ 
$\mathcal{B}_{s,4}$ & ...of which $\rpl < 4~\rearth$  & 342 & 342 \\ 
$\mathcal{B}_{m}$ & Bright ($Kp < 14.2\tablenotemark{b}$) Multis & \nmultip & \nmultis \\ 
$\mathcal{B}_{m,4}$ & ...of which $\rpl < 4~\rearth$ & 415 & 169 \\ 
\enddata
\tablenotetext{a}{Number of transiting planets}
\tablenotetext{b}{Magnitude limit in the \Kepler\ bandpass.}
\end{deluxetable}%

\begin{figure*}[htbp]
\begin{center}
\includegraphics[width=2\columnwidth]{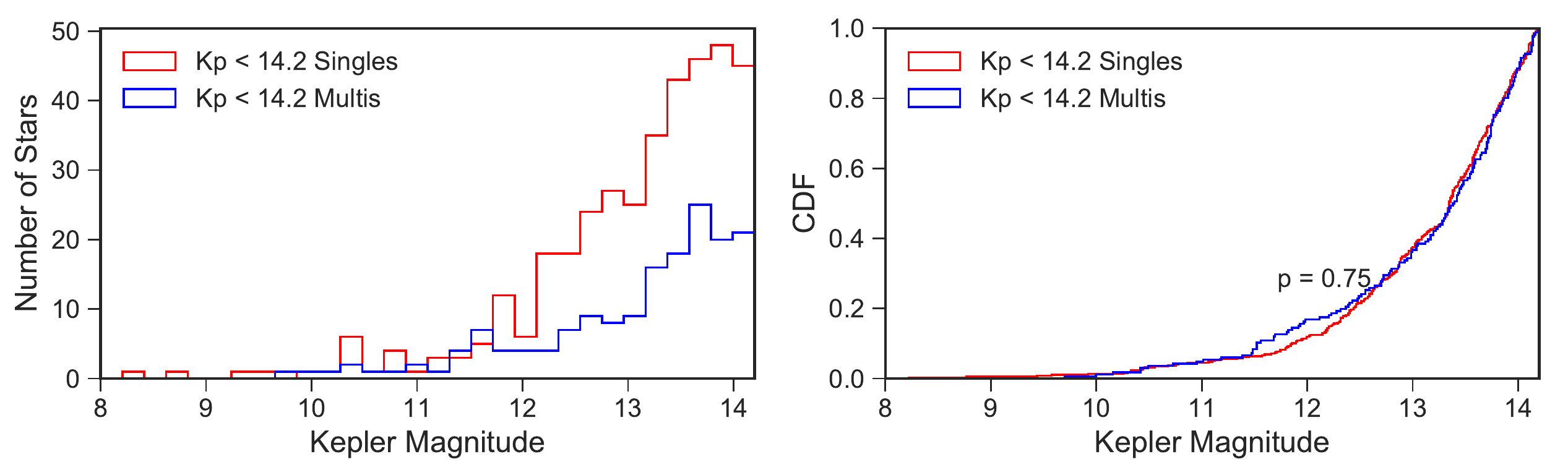}
\includegraphics[width=2\columnwidth]{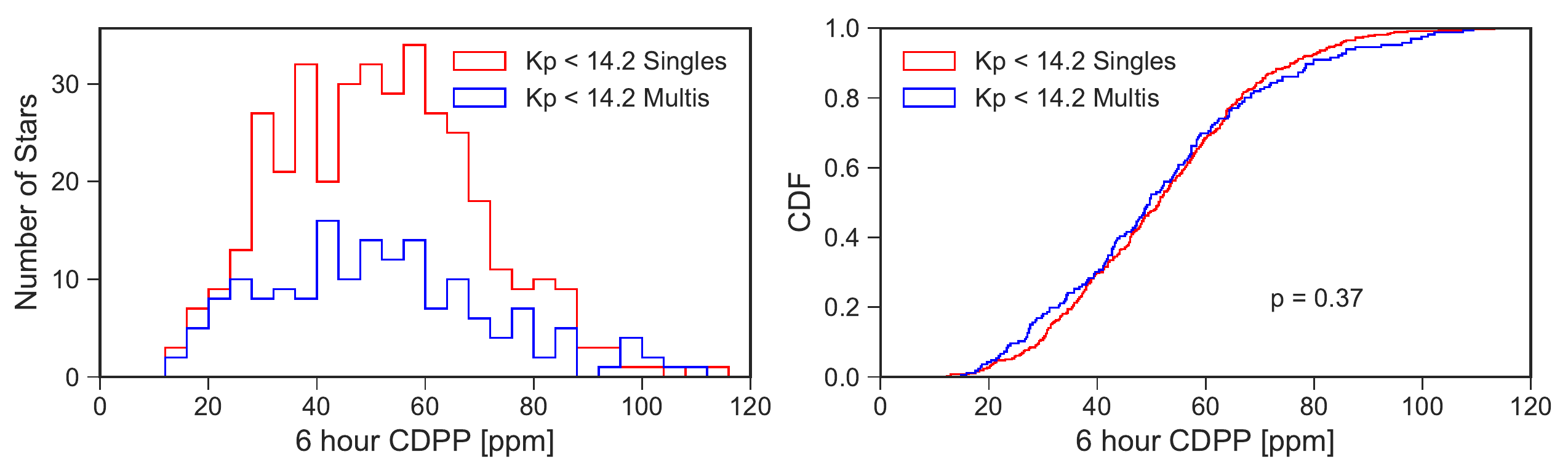}
\caption{Left, top: The host star magnitude in the \Kepler\ bandpass for systems with one transiting planet (red) and multiple transiting planets (blue) in the magnitude-limited CKS samples $\mathcal{B}_s$ and $\mathcal{B}_m$.  Right, top: the same, but the cumulative distribution function (CDF).  An Anderson-Darling test indicates there is no significant distinction between the magnitudes of stars that host one versus multiple transiting planets for $Kp < 14.2$ (p$=0.75$).
Bottom: same as top, but for the combined differential photometric precision (CDPP) in the \Kepler\ bandpass over 6 hour timescales.  There is no significant distinction between the photometric noise of our final samples of singles vs. multis.  These sanity checks show that our selection criteria have not inadvertently favored bright or quiet stars for either the singles or the multis.}
\label{fig:kepmag}
\end{center}
\end{figure*}

As a sanity check, we compare the magnitudes of the host stars of the $Kp < 14.2$ singles and multis.  The distribution of host star magnitudes is indistinguishable for $\mathcal{B}_s$ and $\mathcal{B}_m$, supporting that we have selected samples of singles and multis with similar host star brightnesses.  We also find that the CDPP over 6 hour timescales is indistinguishable for $\mathcal{B}_s$ and $\mathcal{B}_m$.  From the similarity of the CDPP distributions, we conclude that we have not inadvertently selected photometrically noisy stars among either the singles or the multis (see Figure \ref{fig:kepmag}).

\section{Stellar Properties}
\label{sec:stars}
We present the distributions of host star and planetary properties among several samples: $Kp < 14.2$ singles, $Kp < 14.2$ multis, all the CKS-Gaia multis, and the subsets of CKS-Gaia multis with 2, 3, and 4+ transiting planets.  

The stars have effective temperatures from 4500 to 6300 K, masses from 0.5 to 1.6~\msun, radii from to 0.6 to 2.1~\rsun, and projected rotation velocities of $< 20~\kms$.  The uncertainties in stellar effective temperature, mass, radius, metallicity, \vsini, and age are typically 60 K, $0.03~\msun$, $0.03~\rsun$, 0.04 dex, 1~\kms, and 1 Gyr, respectively.

\begin{figure*}[htp]
\begin{center}
\includegraphics[width=2\columnwidth]{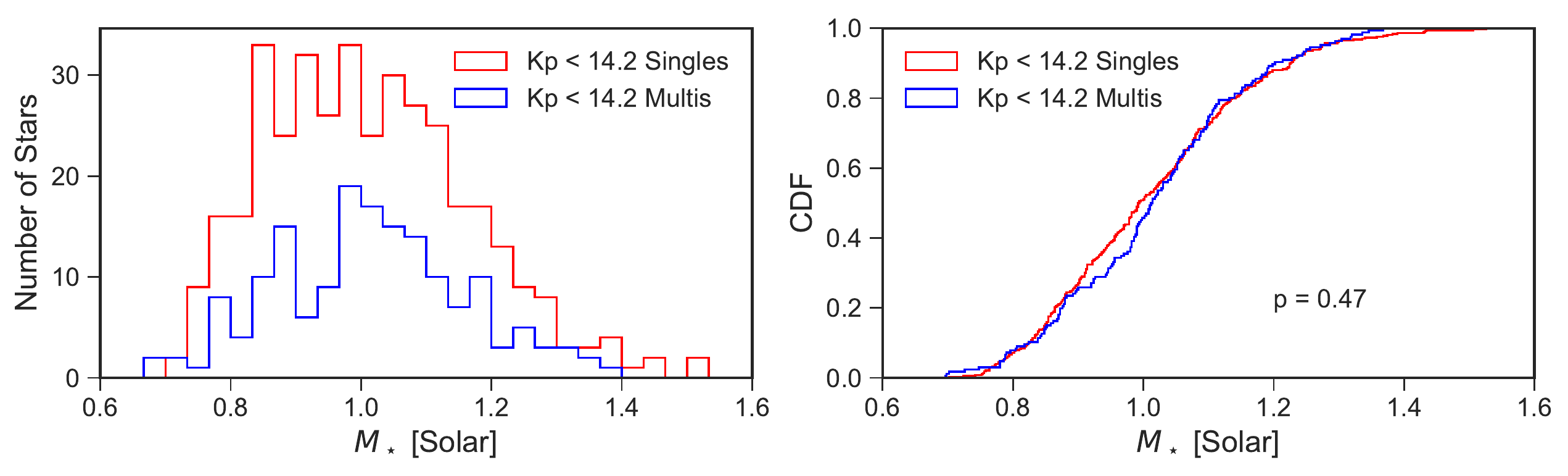}
\includegraphics[width=2\columnwidth]{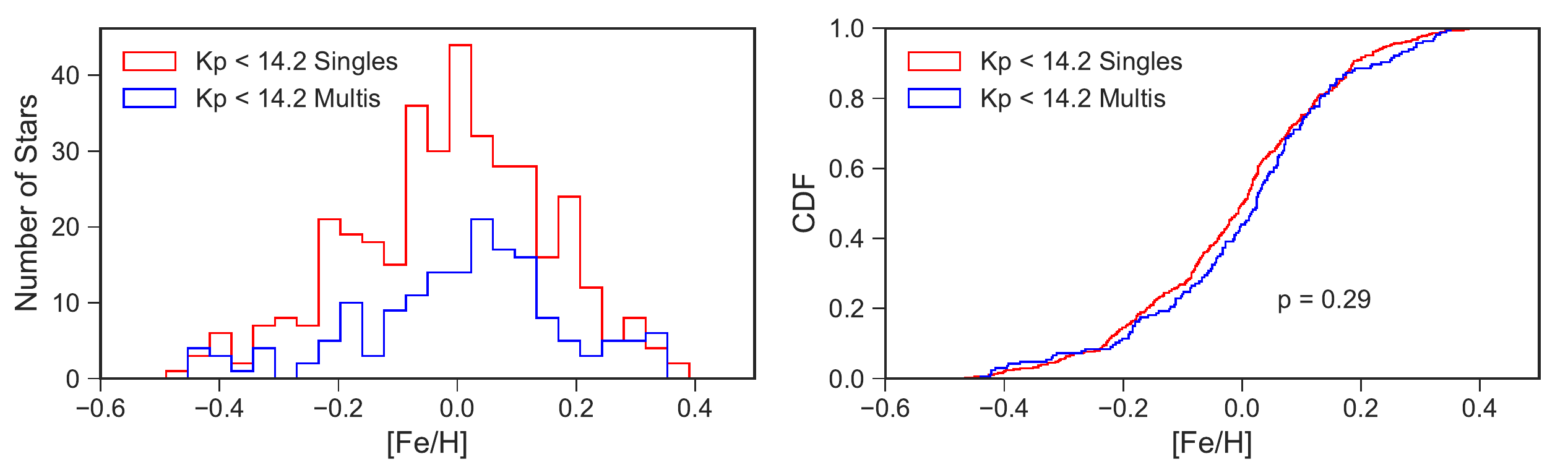}
\includegraphics[width=2\columnwidth]{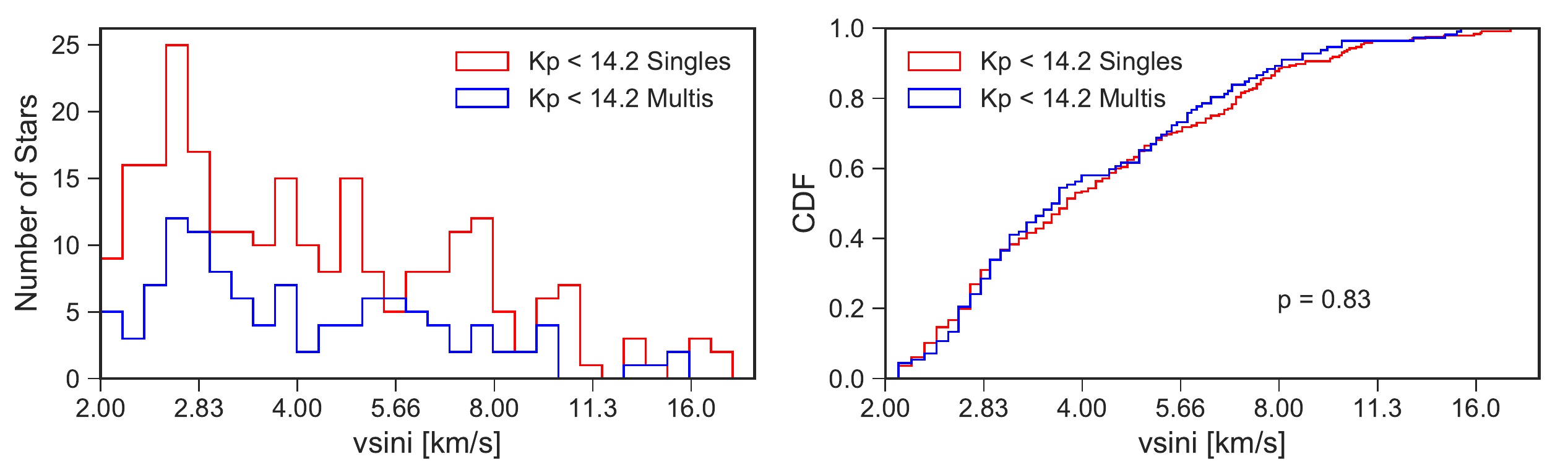}
\caption{Top left: a histogram of the host star masses for systems with one transiting planet (red) and multiple transiting planets (blue) in the magnitude-limited CKS samples $\mathcal{B}_s$ and $\mathcal{B}_m$.  Top right: the same, but the cumulative distribution function (CDF).  The typical uncertainty in stellar mass is $3\%$.  The subsequent rows compare the singles vs. multis distributions for stellar metallicity [Fe/H] (typical uncertainty 0.04 dex), and projected rotation velocity \vsini\ (typical uncertainty $1~\kms$).  With Anderson-Darling tests, we find no significant distinctions between the distributions of singles and multis for any of these stellar parameters ($p > 0.01$).
}
\label{fig:stellar_mvs}
\end{center}
\end{figure*}

We compared the magnitude-limited singles and multis, $\mathcal{B}_s$ and $\mathcal{B}_m$.  Figure \ref{fig:stellar_mvs} shows how the stellar mass, metallicity, and projected rotation velocity distributions of the singles and multis differ.  The panels on the left are histograms of the number of stars; the panels on the right are normalized cumulative distribution functions.  Comparing the distributions of singles and multis with Anderson-Darling tests, we found:
\begin{itemize}
\item no significant difference between the stellar mass (\mstar) distributions of the singles and the multis ($p=0.47$),
\item no significant difference between the stellar metallicity (\feh) distributions of the singles and the multis ($p=0.29$),
\item no significant difference between the projected stellar rotation (\vsini) distributions of the singles and the multis ($p=0.83$).
\end{itemize}
The stellar effective temperatures, radii, and isochrone-determined ages also do not differ significantly.  The host star properties of the $Kp < 14.2$ singles and multis are summarized in Table \ref{tab:mvs}.

In a recent study of the CKS singles and multis, \citet{MunozRomero2018} compared the CKS-determined metallicities of the singles and multis.  Their sample selection removed evolved stars, but did not make the magnitude cuts and detectability cuts we used here.  Nonetheless, they also found no significant differences between the host star metallicities of the CKS singles and multis.

We found no difference in the stellar \vsini\ distributions of the singles vs. multis.  However, the projected rotational velocities are only well-calibrated for $2~\kms < \vsini < 20 ~\kms$ \citep{Petigura2015PhD}.  Thus, the sample of singles and multis with measurable \vsini\ is smaller than the $\mathcal{B}_s$ and $\mathcal{B}_m$ samples: only 245 single stars and 112 multi stars have $2~\kms < \vsini < 20~\kms$.

\begin{figure*}[htp]
\begin{center}
\includegraphics[width=2\columnwidth]{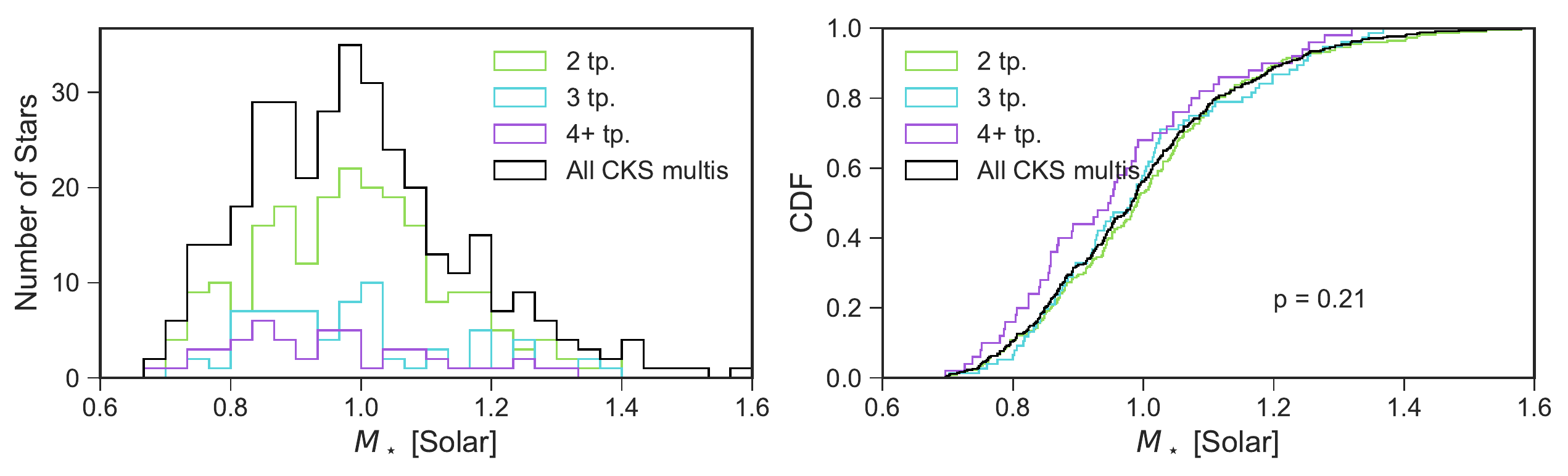}
\includegraphics[width=2\columnwidth]{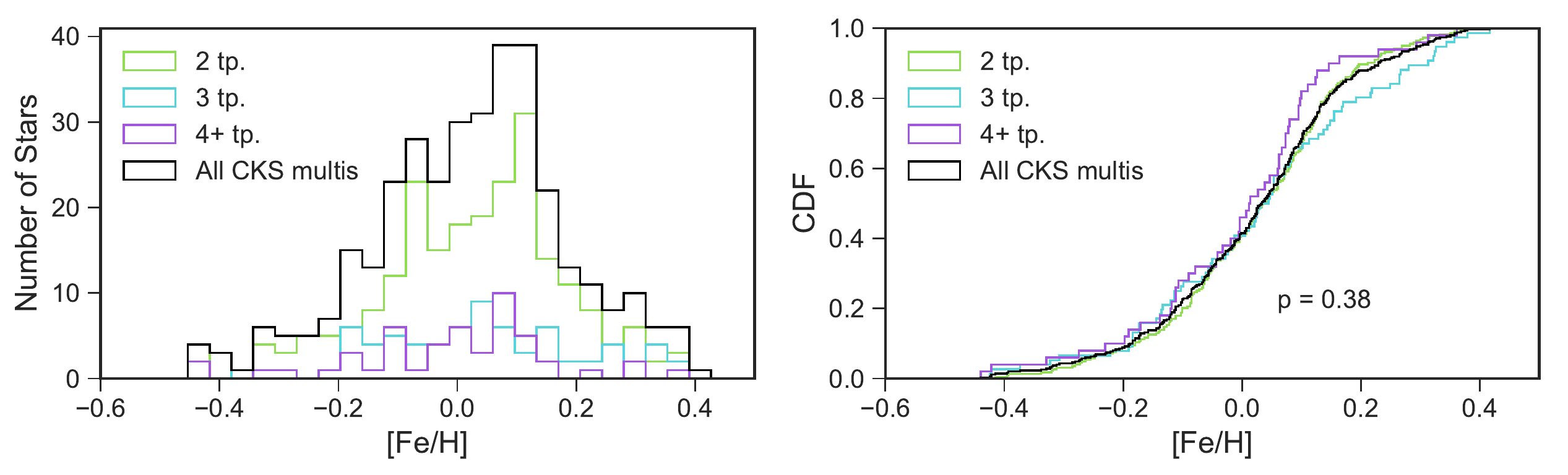}
\includegraphics[width=2\columnwidth]{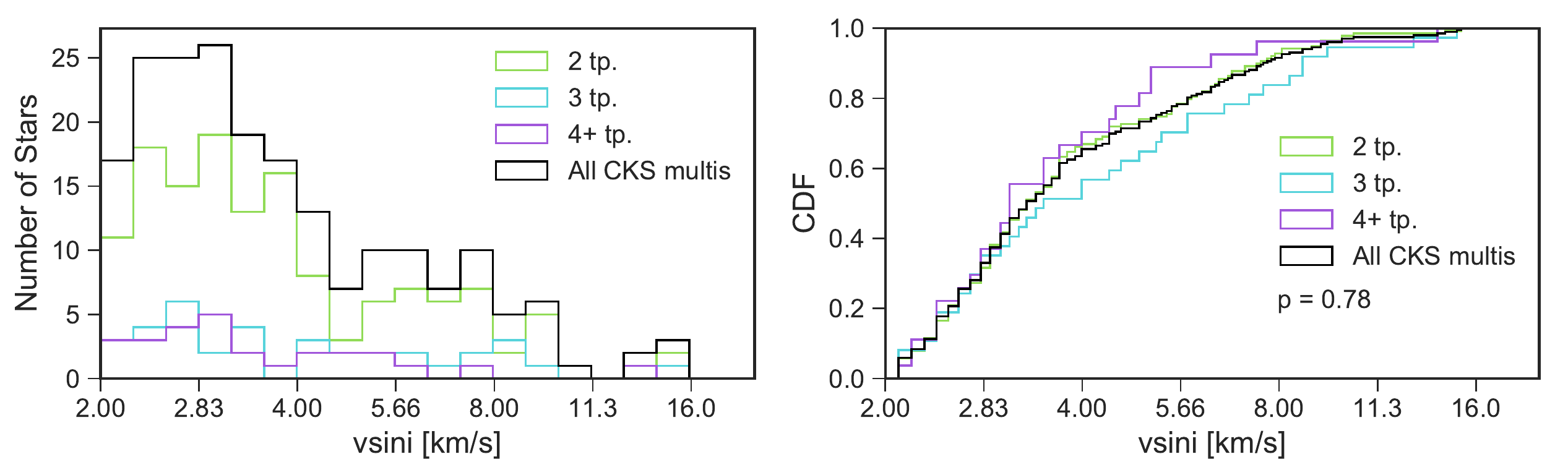}
\caption{The same as Figure \ref{fig:stellar_mvs}, but for the purified CKS-Gaia sample.  The subsets with $\Ntp=2$ (green), $\Ntp=3$ (cyan), and $\Ntp \ge 4$ (violet) are shown.  With Anderson-Darling tests between $\Ntp=2$ and $\Ntp > 2$, we find no significant distinctions between the distributions of stellar properties for the various planet multiplicities.}
\label{fig:stellar_multis}
\end{center}
\end{figure*}

We also compared stars with different transiting planet multiplicities using the full CKS-Gaia sample.  We found no significant differences in the distributions of stellar properties for the systems with 2, 3, and 4+ transiting planets (Figure \ref{fig:stellar_multis}).

The lack of significant correlations between host star properties and planet multiplicities suggests that, if there is some divergence in the planet evolution of the singles vs. multis, such evolution is not dependent on host star type.

\clearpage
\begin{deluxetable*}{lcccccr}
\tablecaption{Statistics of $Kp < 14.2$ Singles vs. Multis \label{tab:mvs}}
\tablewidth{0pt}
\tablehead{
\colhead{Parameter} & \multicolumn{2}{c}{Singles} &  \multicolumn{2}{c}{Multis} & \colhead{Unc.} & \colhead{$p$-value}\tablenotemark{a}\\
\colhead{} & \colhead{Mean} & \colhead{RMS} & \colhead{Mean} & \colhead{RMS} &\colhead{}&\colhead{}}
\startdata
\sidehead{Stellar Properties}
$Kp$ & 13.08 & 0.96 & 13.08 & 0.98 &  & 0.66 \\
CDPP (ppm) & 51.71 & 18.23 & 52.12 & 21.36 &  & 0.36 \\
$R_\star$ ($R_\odot$) & 1.19 & 0.33 & 1.18 & 0.31 & 0.03 & 0.78 \\
Teff [K] & 5737 & 398 & 5733 & 391 & 53 & 0.70\\
$M_\star$ ($M_\odot$) & 1.01 & 0.15 & 1.02 & 0.15 & 0.03 & 0.43 \\
\feh & -0.01 & 0.17 & 0.01 & 0.17 & 0.04 & 0.22 \\
\vsini~ (km/s)\tablenotemark{b} & 4.86 & 3.03 & 4.68 & 2.74 & 1.00 & 0.86 \\
Age (Gyr) & 5.69 & 3.38 & 5.51 & 3.27 & 1.54 & 0.68 \\
\hline
\sidehead{Planet Properties}
$\rpl$ & 2.51 & 2.74 & 2.03 & 1.42 & 0.09 & 0.02 \\
Per (days) & 35.97 & 89.60 & 28.14 & 67.39 & 0.00021 & 0.001 \\
\sidehead{Planet Properties ($\rpl < 4~\rearth$)}
$\rpl$ & 1.76 & 0.79 & 1.80 & 0.73 & 0.09 & 0.072\\
Per (days)	 & 22.27 & 42.49 & 23.01 & 39.65 & 0.00021 & 0.002\\
\sidehead{Planet Properties ($\rpl < 4~\rearth$ \& $P > 3$ days)}
$\rpl$ & 1.86 & 0.78 & 1.84 & 0.73 & 0.09 & 0.44 \\
Per (days) & 26.11 & 45.29 & 24.47 & 40.60 & 0.00023 & 0.32 \\
\enddata
\tablenotetext{a}{Anderson-Darling test $p$-value}
\tablenotetext{b}{For $\vsini > 2 ~\kms$}
\end{deluxetable*}

\section{Planet Properties}
\label{sec:planets}
\begin{figure*}[htbp]
\begin{center}
\includegraphics[width=2\columnwidth]{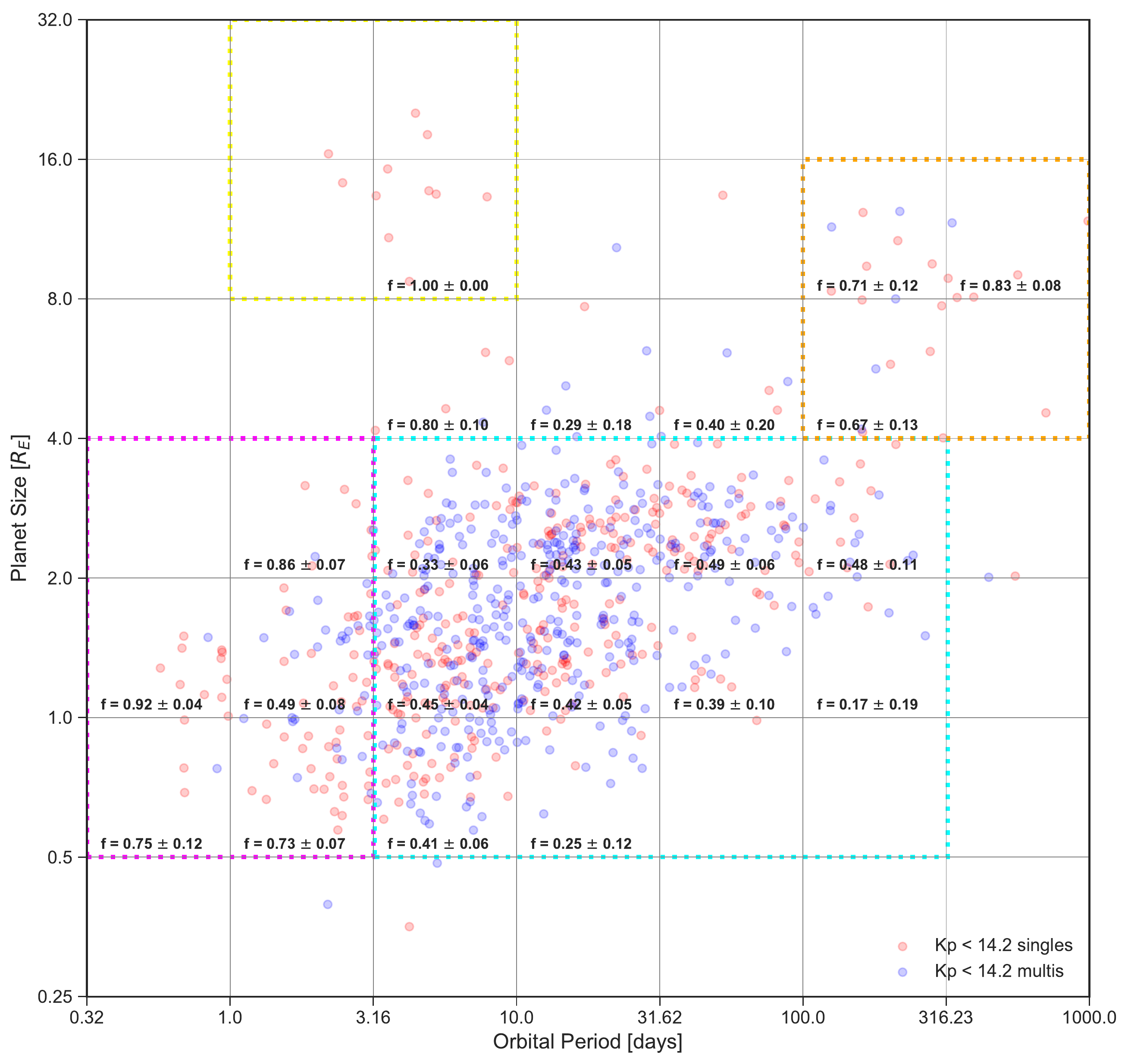}
\caption{The CKS singles (red) and multis (blue) as a function of orbital period and planet size.  The fraction of detected planets that are in singles, $f$, is displayed in each grid cell with at least 4 planets.  The uncertainties are calculated using binomial statistics.  The fraction of detected sub-Neptunes with $P > 3$ days (i.e., the majority of the the \Kepler\ planets) that are singles is $43\pm2\%$ (cyan box).  However, among the detected sub-Neptunes with $P < 3 $ days, $68\pm5\%$ are singles (magenta box).  The 11 hot Jupiters (yellow box) are all singles.  The population of cold giants (orange box) is also predominantly singles ($71\pm9\%$).  We explore each of these regions of parameter space in the text.}
\label{fig:rp_v_per}
\end{center}
\end{figure*}

We examined the distributions of singles and multis in the planet radius-orbital period plane (Figure \ref{fig:rp_v_per}).  The fraction of detected planets that are singles, $f$, in each grid cell is given.  The uncertainty in $f$ is the 68\% confidence interval calculated from binomial statistics.  Note that because our definitions of singles and multis are based on the transiting planet multiplicity, the values of $f$ and their uncertainties do not necessarily describe the true planet multiplicity.  Rather, $f$ is an observed quantity that should be reproduced by future attempts to model underlying planetary architectures.  In this section, we identify regions in period-radius space where contiguous cells have similar values of $f$ and discuss possible interpretations.  We follow with a discussion of the the 1-D distributions of planet radius and orbital period for the singles and multis.

\subsection{Sub-Neptunes with $P > 3$ days}
Sub-Neptune sized planets ($\rpl < 4 \rearth$) are the majority of the planets in the CKS-Gaia sample, and they are also intrinsically common \citep{Petigura2013,Fressin2013}.  Most sub-Neptunes have $P > 3$ days (we will explore the population with $P < 3$ days below).  Figure \ref{fig:rp_v_per} shows the sub-Neptunes with $P > 3$ days and $0.5~\rearth < \rpl < 4~\rearth$ in a cyan box.  Within the cyan box, a detected planet is slightly less likely to be a single than a multi ($f = 43\pm2\%$).  However, there is very little variation in the fraction of singles as a function of period and radius within the cyan box.  Thus, for the majority of the \Kepler\ planets, the orbital period and size of the planet are not good predictors of whether the planet will have additional transiting companions.

The sub-Neptunes dominate the sample, hence their host star properties are very similar to the host star properties of the sample overall.  The mean mass of the sub-Neptunes with $P > 3$ days is 1.01~\msun for the singles (1.01 for the multis), and the mean value of [Fe/H] is -0.03 for the singles (-0.02 for the multis).

\subsection{Hot Jupiters}
The hot Jupiters ($P < 10$ days, $8~\rearth < \rpl < 22~\rearth$, yellow box in Figure \ref{fig:rp_v_per}) in the CKS-Gaia sample are all singles.  The high fraction of singles among the hot Jupiters confirms the well-studied phenomenon that hot Jupiters are lonely \citep[e.g., ][]{Steffen2012}.  Recall that in our sample of hot Jupiters, we would have been able to detect any transiting planets with $\rpl > \rthresh$ and $P < \Pthresh$, based on our selection critera.  If the next hot Jupiter discovered had transiting companions, the hot Jupiters would have $f=92\pm6\%$.  Combining this fraction with the overall occurrence of hot Jupiters ($\sim1\%$), hot Jupiters with nearby coplanar companions occur around no more than $\sim0.1\%$ of stars.  Thus, WASP-47, the only known hot Jupiter with small transiting companions \citep{Becker2015}, belongs to a rare population.

Hot Jupiter host stars have higher masses and metallicities than field stars \citep{Johnson2007, Fischer2005}.  We compare the host star properties of the hot Jupiters to the host star properties of the small exoplanets ($\rpl < 4~\rearth$).  The hot Jupiters orbit more massive and more metal-rich stars than the sub-Neptunes.  The mean mass and metallicity of a hot Jupiter host star are $\mstar = 1.17~\msun$ and $\mathrm{\feh} = 0.16$, whereas the mean mass and metallicity for a sub-Neptune's host star are  $\mstar = 1.01~\msun$ and $\mathrm{\feh} = 0.00$.  The typical uncertainties in the stellar masses and metallicities are 3\% and 0.04 dex, respectively, and Anderson-Darling tests yield with $>99\%$ confidence ($p < 0.01$) that the host star masses and metallicities of the hot Jupiters are drawn from different distributions than the host star masses and metallicities of the sub-Neptunes.

\subsection{Cold Jupiters}
Cold Jupiters are giant planet candidates with $P > 100$ days, $5~\rearth < \rpl < 16~\rearth$ (orange box in Figure \ref{fig:rp_v_per}).  The fraction of singles among the cold Jupiters is high compared to the sub-Neptunes ($f = 71\pm9\%$), but not as high as the hot Jupiters.  Formally, the excess of giant cold singles (compared to the near-parity of sub-Neptune sized singles and multis) is significant with $3\sigma$ confidence, but concluding that most of the cold giant planets are indeed singles is premature.  Not all of the long-period giant planet candidates are confirmed, and false positives are common for planets of these sizes and orbital periods \citep[e.g., as many as 35-50\% of the unconfirmed giant planets could be false positives,][]{Santerne2016}.  We have already excluded known and likely false positives from the CKS sample.  However, only 9 out of 17 cold giant singles are already confirmed.   \citep[The giant planets in multis are statistically validated, e.g.][]{Lissauer2012}.  Vetting the remainder of the single, cold giant planet candidates would clarify whether the excess of single cold giants (compared to sub-Neptunes) is real.  

The host star properties of the single, cold giant planet candidates do not differ substantially from the host star properties of the sub-Neptunes (Anderson-Darling $p > 0.1$ in comparisons of stellar mass, metallicity, and \vsini).  The average mass of the cold Jupiter host stars is 1.07 \msun, which is slightly higher than the typical stellar mass for the sub-Neptunes (1.0 \msun), but lower than the typical host-star mass for hot Jupiters (1.17 \msun).  The average metallicity of the cold giant planet host stars is [Fe/H] $= 0.01$, which does not differ significantly from the average metallicity of the sub-Neptune host stars ([Fe/H] $= 0.0$).

\subsection{Hot Sub-Neptunes}
In particular, there is an excess of singles relative to multis among the hot sub-Neptunes ($\rpl < 4~\rearth$, $P < 3$ days, magenta box in Figure \ref{fig:rp_v_per}).  There are 80 planets in this size and period range, of which 54 are singles ($f=68\pm5\%$), resulting in a significantly higher fraction of singles than for the sub-Neptunes with $P > 3$ days ($f = 43\pm2\%$).

The vast majority (74/80) of the sub-Neptunes with $P < 3$ days have $\rpl < 1.8~\rearth$.  There is mounting evidence that planets smaller than 1.8~\rearth\ and close to their stars are rocky: the masses of many planets with $\rpl < 1.5~\rearth$ and $P < 100 $ days are consistent with rocky compositions \citep{Weiss2014, Rogers2015}.  Furthermore, there are two distinct size populations of small planets, with a valley in the planet radius distribution at $1.8~\rearth$ \citep{Fulton2017}.  Planets smaller than 1.8~\rearth\ and near their stars are likely the rocky cores of photo-evaporated planets \citep{Fulton2017, Owen2017}.  Hence, the planets with $P < 3$ days and $\rpl < 1.8~\rearth$ might better be described as ``hot super-Earths'' than sub-Neptunes.

The host star properties of the hot sub-Neptunes are very similar to the host star properties of the sub-Neptunes with $P > 3$ days.  The metallicities are slightly higher on average ($\mathrm{\feh = 0.01}$) for the hot sub-Neptunes than the cool sub-Neputunes ($\mathrm{\feh = -0.03}$).  However, an Anderson-Darling test yields $p > 0.1$, indicating that the distribution of host star metallicities for the hot super-Earths is consistent with the distribution of host star metallicities for the cool sub-Neptunes.  \citet{Petigura2018} also found slightly higher host star metallicities for the hot super-Earths than for the cool sub-Neptunes.  However, the scatter in host star metallicities for both the singles and multis (RMS=0.17 dex) is much larger than the slight difference in their average host star metallicities (0.04 dex).

\citet{Steffen2016} also identified an excess of detected short-period, Earth-sized singles compared to multis.  That study used a kernel estimation to identify clusters of singles and multis in ($\rpl$, $P$) space, whereas we have divided ($\rpl$, $P$) space into a grid.  Our study confirms the population of detected short-period planets that are predominantly singles.

\subsection{Planet Radius}
\begin{figure*}[htbp]
\begin{center}
\includegraphics[width=2\columnwidth]{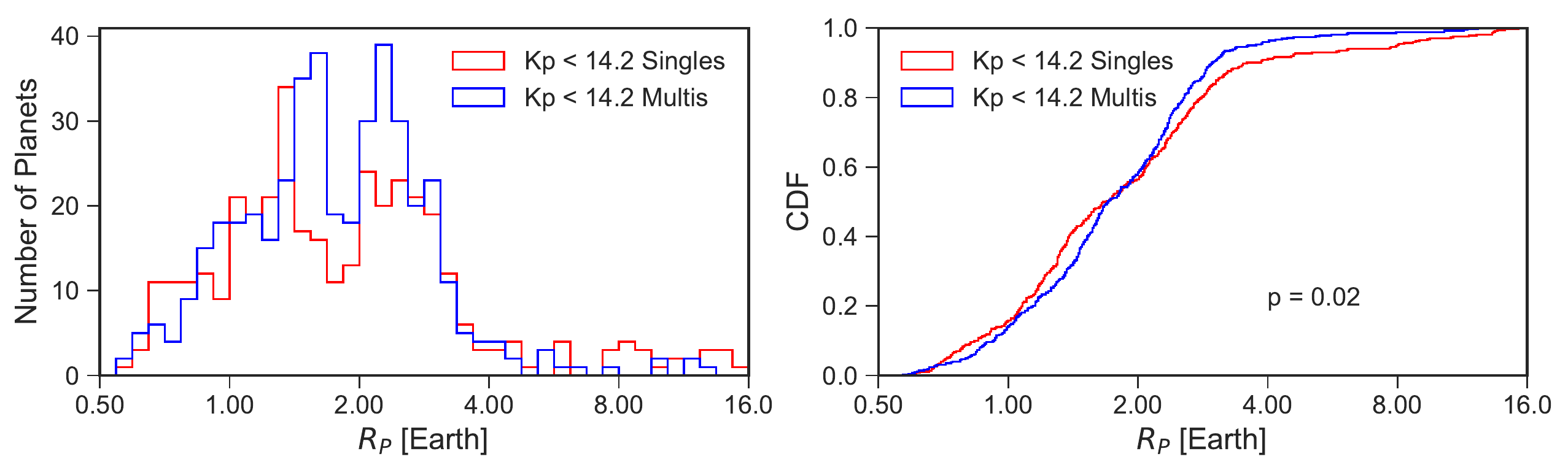}
\includegraphics[width=2\columnwidth]{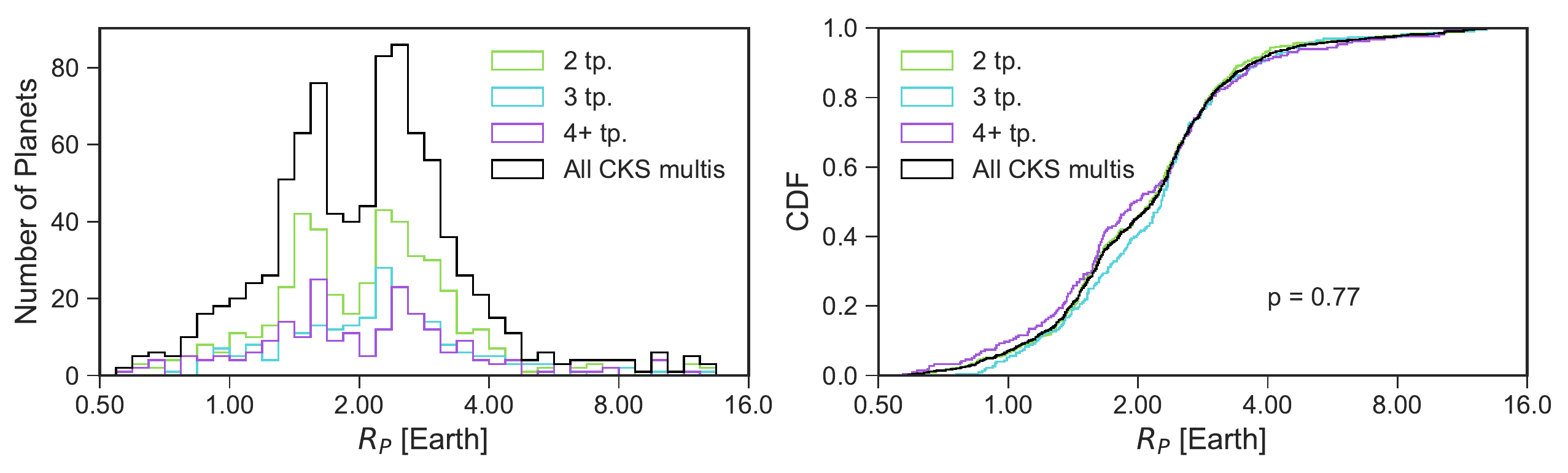}
\includegraphics[width=2\columnwidth]{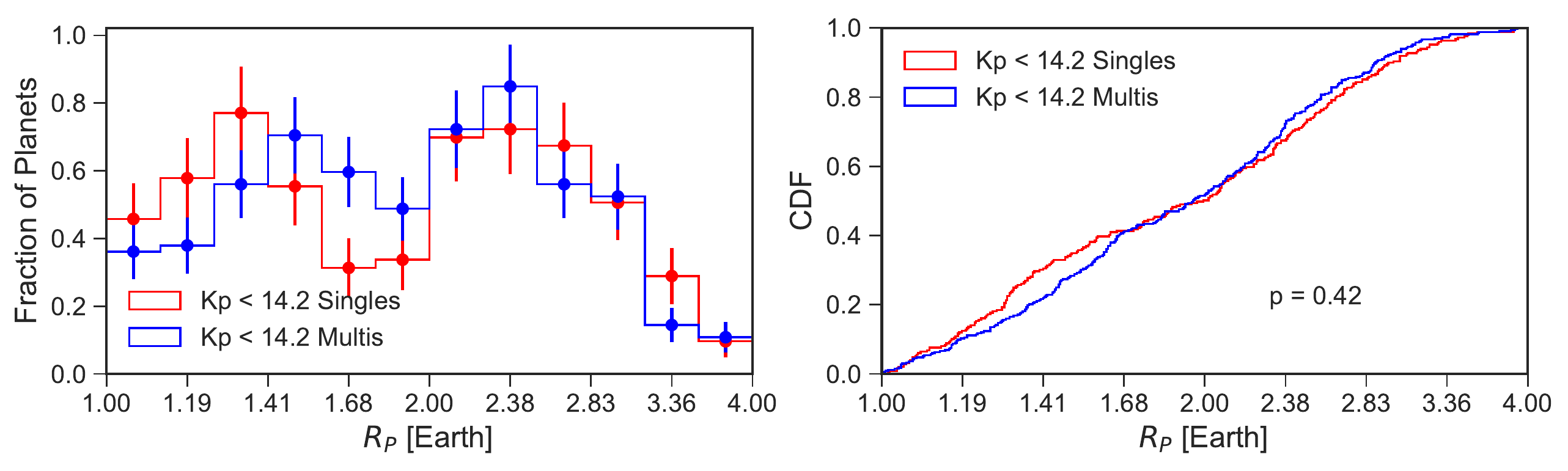}
\caption{Top panel, left: the distribution of planet radius for the CKS $Kp < 14.2$ systems with one (red) and multiple (blue) transiting planets. Top panel, right: the same, but the cumulative distribution function (CDF).  The majority of planets in singles and multis are smaller than $4~\rearth$, and both distributions show a valley at $1.8~\rearth$.  The tail of the singles distribution includes more giant planets than the multis. Middle panel: same as the top panels, but for all of the CKS-Gaia multis (black), and for the sub-samples with 2, 3, and 4+ transiting planets, all of which are consistent with a valley at $1.8~\rearth$.  Bottom panel: zoom of the valley in singles and multis for $1~\rearth < \rpl < 4~\rearth$ and $P > 3$ days, including Poisson errors.  The prevalence of the valley at 1.8~\rearth\ in the CKS-Gaia singles and multis indicates that the dynamical history that makes singles ``singles'' is unrelated to the formation of the radius valley.}
\label{fig:rp}
\end{center}
\end{figure*}
The top panel of Figure \ref{fig:rp} shows the distribution of planet radii for $\mathcal{B}_{s}$ and $\mathcal{B}_{m}$.  There is a valley in planet radius at 1.8~\rearth\ in both the singles and multis.  This valley was announced in \citet{Fulton2017} (CKSIII), but that paper did not specifically address whether the valley exists in multi-planet systems.  Here, we find that multi-planet systems indeed have a valley in the distribution of planet radii at about 1.8~\rearth.  

The middle panel of Figure \ref{fig:rp} shows the distribution of planet sizes for planets that belong to systems with 2, 3, and 4+ transiting planets.  The radius valley at 1.8~\rearth\ is evident in the complete CKS-Gaia sample (black), which includes multis having $Kp > 14.2$.  This large sample (\npl transiting planets) clarifies the existence of the radius valley.  The sub-samples with $\Ntp=2$ (green), $\Ntp=3$ (cyan), $\Ntp \ge 4$ (purple) are all consistent with a valley, although the valley is less clear in the high-multiplicity systems simply because there are fewer planets.

As discussed above, there is an excess of single giant planet candidates.  An Anderson-Darling test comparing the all of the planet radii in $\mathcal{B}_{s}$ and $\mathcal{B}_{m}$ yields a $p$-value 0.02 (Figure \ref{fig:rp}).  Focusing on the sub-Neptune sized planets with $P > 3$ days only (cyan box in Figure \ref{fig:rp_v_per}), the $p$-value is 0.44, indicating no significant difference between the radius distributions of the singles and multis (bottom panel of Figure \ref{fig:rp}).  The CDFs of the sub-Neptune sized singles and multis have the greatest differences near the valley at 1.8~\rearth, but the differences in their distributions are not statistically significant.  This is because there are not enough sub-Neptune sized planets near the valley to determine whether the shape or position of the valley differs between the singles and multis.  Nonetheless, we detect the presence of the valley in the magnitude-limited singles and multis ($\mathcal{B}_{s,4}$ and $\mathcal{B}_{m,4}$) as well as in the full sample of CKS-Gaia.  

The radius valley exists for the planets in singles and multis.  If some physical process (such as planet-planet scattering) disrupts multi-planet systems to make the singles, that process does not erase or significantly alter the radius valley.  The radius valley is thought to be sculpted within the first 100 Myr of the planetary system's lifetime.  Photo-evaporation, which scales with the inverse square of orbital distance, is likely responsible at least in part for the presence of the radius valley \citep{Owen2017}, although other mechanisms to strip a planet's volatile envelope have been proposed \citep{Ginzburg2018}. 

The similar radius distributions thus suggest that the singles and multis have similar migration histories (or lack thereof), and that any large-scale migration likely happened in the first 100 Myr, bringing the planets close enough to the stars for photo-evaporation to produce the radius valley.  If the planets of one population (say, the singles) had migrated great distances after photo-evaporation turned off (i.e., after 100 Myr), we would have seen less of a gap in that population, as photo-evaporation is weak at large distances and at late times.

\subsection{Planet orbital period}
\begin{figure*}[htbp]
\begin{center}
\includegraphics[width=2\columnwidth]{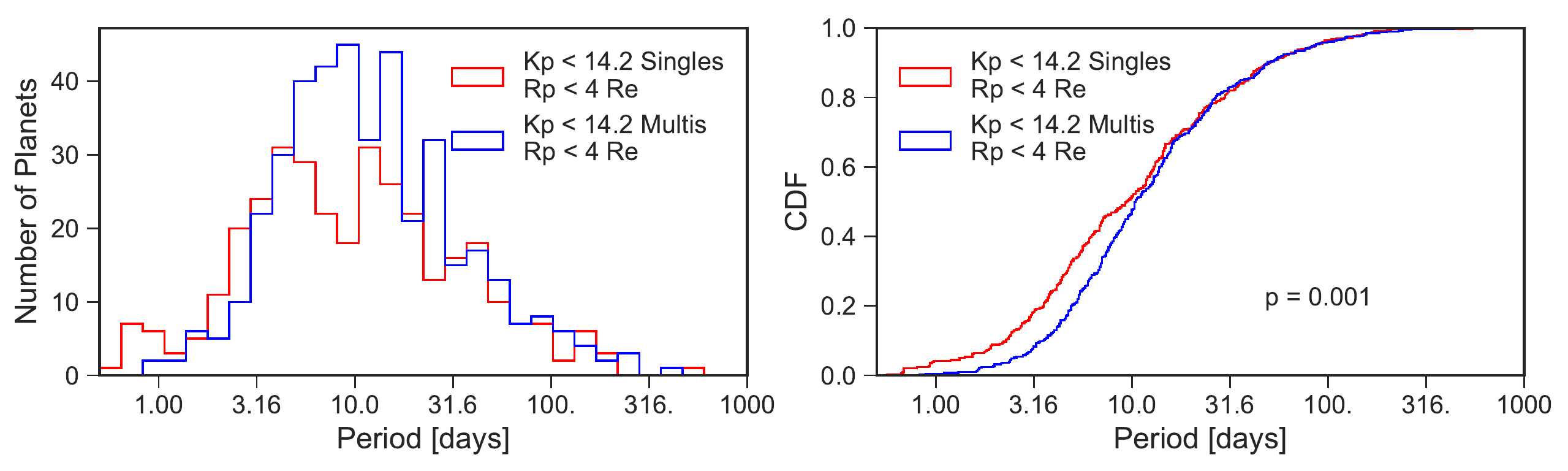}
\includegraphics[width=2\columnwidth]{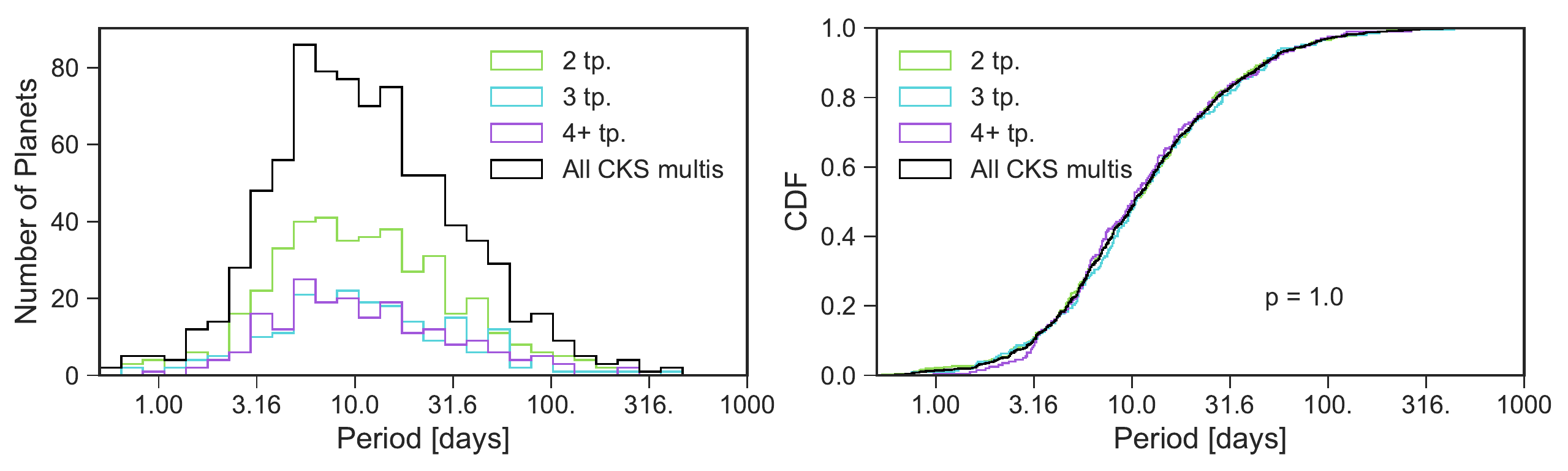}
\caption{Top left: the distribution of orbital periods for the CKS $Kp < 14.2$ sub-Neptunes that are singles ($\mathcal{B}_{s,4}$, red) and multis ($\mathcal{B}_{m,4}$, blue). Top right: the same, but the cumulative distribution function (CDF).  There is an excess of short-period singles ($P < 3$ days) among the sub-Neptune sized planets. Bottom left: the distribution of planet orbital period for all of the CKS-Gaia (black), and for the sub-samples with 2, 3, and 4+ transiting planets. Bottom right: the CDF.  There are no significant differences in the distributions of orbital periods based on transiting planet multiplicity.  If the excess of singles at $P < 3$ days were due to geometric effects, we would likely also see an excess of systems with 2 transiting planets as compared to 3, and 3 transiting planets as compared to 4+, due to geometric effects.  The similarity of the orbital period distributions of multis with different transiting planet multiplicities suggests that some difference in orbital architecture might account for the excess of short-period singles.}
\label{fig:per}
\end{center}
\end{figure*}

The top panels of figure \ref{fig:per} show the distribution of planet orbital periods for the sub-Neptunes in $\mathcal{B}_{s,4}$ and $\mathcal{B}_{m,4}$.  There are more singles at short orbital periods ($P < 3$ days) than multis.  Beyond 10 days, the orbital periods of the singles and multis are similar.  An Anderson-Darling comparison of the period distributions of singles vs. multis yields a $p$-value of 0.001, indicating that these distributions are not drawn from the same underlying distribution with $>99\%$ confidence.  

Planets at short orbital periods are more likely to transit than planets at long orbital periods.  Could the excess of short period singles be the result of a geometrical viewing effect that has nothing to do with astrophysics?   If geometry alone is responsible for the excess of short period singles, we might expect to see a difference between the $\Ntp =2$, 3, and 4+ sub-samples in this period range.  However, the bottom half of Figure \ref{fig:per} shows that the orbital distributions for multiple transiting planets are indistinguishable for $\Ntp=2$, 3, and 4+.  In other words, there are just as many planets with $P < 3$ days that belong to 4-transiting planet systems as there are planets that belong to 2 or 3-transiting planet systems.  That the various multi-planet systems have very similar orbital period distributions, whereas the distribution of orbital periods of singles is unique, suggests that geometrical bias alone is unlikely to account for the excess of singles, although a detailed suite of forward-modeling is necessary to demonstrate this claim.

In principle, the excess of singles at short orbital periods could possibly be related to more frequent false positives at short periods.  However, false positives are unlikely to mimic such a large number of planets considering the high purity of our sample.

Alternatively, the hot super-Earths with $P < 3$ days might have a different dynamical history from the longer-period sub-Neptunes, as suggested in \citet{Steffen2016}.  The ultra-short periods planets ($P < 1$ day) with additional transiting planets have wider period ratios between the innermost pair of planets than the farther-out planet pairs \citep{Sanchis-Ojeda2014}.  In contrast, the majority of transiting planets in multi-planet systems have very regular orbital period spacing \citep{Weiss2018}.  Thus, it is likely that a dynamical process is responsible for the inward migration of the innermost planet in multi-planet systems.  Planet-planet scattering \citep[e.g.,][]{Chatterjee2008}, secular chaos \citep{Petrovich2018}, and tidal inspiral \citep{Lee2017} are theories that systematically bring the innermost planet of a multi-planet system to short orbital periods.

Why do the inward-moved small planets in our sample tend to be singles?  The large period ratio between the innermost planet and the next planet out might be enough to explain why the short-period planets tend to be singles.  Suppose the short-period singles really belong to nearly-coplanar multi-planet systems whose midplanes are slightly misaligned relative to our line of sight.  In this case, the combination of the orbital distance and inclination of the innermost planet might allow it to transit, while the other planets are too distant from the star to transit.  However, this explanation is inconsistent with the lack of observed short-period planets in multis.  If a dynamical process were frequently moving the innermost planet closer to the star \textit{without} disrupting the coplanarity of the system, we would expect to frequently detect the short-period ($P < 3$ days) sub-Neptunes among the multis.  Therefore, in addition to a mechanism that moves the innermost planet closer to the star, a mechanism that increases the mutual inclinations of the planets is likely needed to explain why there are so many short-period singles and so few short-period multis among the sub-Neptunes.  Recently, \citet{Dai2018} found that the mutual inclinations of planets in multis with a very short-period planet are larger than the mutual inclinations in multis with no transiting short-period planets.

Ultimately, the orbital period distributions of the $\mathcal{B}_s$ singles, $\mathcal{B}_m$ multis, and the $\Ntp=2$, 3, and 4+ samples are all empirical distributions that should be reproduced by any model that aims to describe the underlying distributions of the planet multiplicities, orbital periods, and inclinations.  \citet{Fang2012,Tremaine2012,Ballard2016,Gaidos2016} and \citet{Zhu2018} all attempted to determine the underlying multiplicity and inclination distributions of the \Kepler\ multis, but none of these attempts sought to reproduce the orbital period distributions for the different transiting planet multiplicities.  Such an exercise would be valuable but is beyond the scope of this paper.  A new tool presented in \citet{Mulders2018} is a promising step toward self-consistently modeling the various observed distributions of the singles and multis.


\section{Conclusions}
\label{sec:conclusion}
In this paper, we explored how the physical properties of the CKS systems containing multiple detected transiting planets (multis) compare to systems with just one detected transiting planet (singles).  Although other studies have examined the relationships between stellar and/or planetary properties and planet multiplicity, our study presents three advantages: (1) The CKS-Gaia dataset enables the largest, most accurate, and most precise comparison of the fundamental host star properties of the \Kepler\ singles and multis so far.  (2) As a result of stringent magnitude, detection threshold, and false positive cuts, our comparison of singles vs. multis suffers from fewer observational biases than other studies.  (3) In addition to comparing the properties of singles vs. multis, we compare the host star and planet properties as a function of the number of transiting planets.

Our conclusions are as follows:
\begin{enumerate}
\item \textit{The distributions of stellar mass, metallicity, and projected rotation velocity do not differ significantly for the singles and multis.}  The lack of a relationship between stellar physical properties and apparent planet multiplicity suggests that any physical process that preferentially creates ``singles'' occurs late in planet formation and in a manner that is not related to the properties of the host star.  Also, stellar properties are not particularly useful in predicting the number of transiting planets around a star.  
\item \textit{Transiting planets of various multiplicities exhibit a valley in the radius distribution at $\sim1.8~\rearth$}.  The statistically indistinguishable size distributions of small planets ($\rpl < 4\rearth$) in singles and multis suggests that the acquisition of and subsequent evaporation of a volatile envelope around the planetary core is the same for the singles and multis.  Because photo-evaporation happens within the first 100 Myr and is only effective at short orbital periods, the singles and multis likely arrive near their present orbital distances within the first 100 Myr.
\item \textit{For the sub-Neptune sized planets, there is a  significant ($p = 0.001$) excess of short-period singles ($P < 3$ days) compared to multis.}  However, among the multis, the $\Ntp = 2$, 3, and 4+ orbital period distributions are the same, suggesting that geometrical bias alone is unlikely to explain the excess of short-period singles.  False positives are unlikely to mimic such a large number of planets considering the high purity of our sample.  The excess of short-period planets could also be the hallmark of a late mechanism, such as planet-planet scattering or tidal migration, that produces more singles than multis at very short orbital periods.
\item Hot Jupiters are almost always single transiting planets that orbit high-mass and high-metallicity host stars, but these systems are intrinsically rare.
\end{enumerate}

Our main finding is that host star properties and planet radii for the majority of the sub-Neptune sized planets have no strong relationship with whether there are multiple transiting planets in the system.  The similarity of the singles and multis suggests that they have a common origin.  The majority of the singles with $P > 3$ days and $\rpl < 4 \rearth$ likely belong to multi-planet systems with higher mutual inclinations than the CKS multis.  Perhaps the singles are multi-planet systems that have undergone planet-planet scattering, resulting in systems of multiple planets in which only one planet transits.  By contrast, the multis likely had dynamically quieter histories, as evidenced by their current low-entropy states.

How might planet composition relate to multiplicity?  At face value, the observed radii and orbital periods of the singles and multis are inconsistent with the prediction in \citet{Dawson2016}.  That study predicted that planets in multi-planet systems should have preferentially volatile-rich compositions, whereas the singles should preferentially have rocky compositions.  The idea behind the prediction was that the multi-planet systems form a little bit earlier than the singles, while the gas disk is a little bit denser, which contributes to both (1) eccentricity damping, resulting in more circular (hence, stable) orbits for the planets, and (2) more gas-rich planets.  In contrast, we observe that both the singles and multis include significant populations of planets larger and smaller than the transition from rocky to volatile-rich planets at $\sim1.8~\rearth$.  In other words, there is no evidence that the singles are preferentially rocky, or that the multis are preferentially gas-rich.  Furthermore, if there were an especially large population of rocky singles with $10 < P < 30$ days and $1~\rearth < \rpl < 1.8~\rearth$, many such planets would have been detected in the \Kepler\ Mission and included in our sample.

Perhaps the singles and multis generally form in compact multi-planet systems.  These planets can either form early while gas is abundant (forming volatile-rich planets) or later when there is less gas (forming rocky planets).  In either case, whatever small eccentricities the planets have acquired will grow after the gas disk dissipates, sometimes leading to dynamical instability.  For instance, \citet{Obertas2017} found that the Lyapunov time can vary by a couple orders of magnitude for compact multi-planet systems based on slight differences in the initial orbital conditions\footnote{Although it might not be obvious how the initial conditions of some compact multi-planet systems (but not others) lead to eventual instability, \citet{Tamayo2018} found that machine learning techniques are able to predict the outcomes of N-body simulations with high fidelity.}. 
Thus, whether a multi-planet system becomes unstable on a timescale of gigayears might be primarily assigned at birth.  On the other hand, external influences such as passing stars might also play a role in dynamically disrupting initially coplanar multis \citep{Spalding2016}.  The singles have higher eccentricities on average than the planets in multis \citep{Xie2016, vanEylen2018_ecc}, suggesting that dynamical heating and perhaps instability play a role in the formation of the singles.

Single sub-Neptunes at very short orbital periods likely have a different dynamical history than the multis.  The singles with $P < 3$ days likely belong to multi-planet systems in which some combination of planet-planet scattering, secular chaos, and/or tidal inspiral has moved the innermost planet close to its star.  Additional measurements, especially of the impact parameters, orbital obliquities, eccentricities, and masses of the planets in both singles and multis with $P < 3$ days, will clarify which astrophysical process best explains the apparent excess of sub-Neptune sized singles at short orbital periods.

\acknowledgments
We thank James Owen, Hilke Schlichting, Daniel Fabrycky, and the anonymous referee for helpful conversations and feedback.
Most of the data presented here were determined directly from observations at the W.\ M.\ Keck Observatory, which is operated as a scientific partnership among the California Institute of Technology, the University of California, and NASA. We are grateful to the time assignment committees of the University of Hawaii, the University of California, the California Institute of Technology, and NASA for their generous allocations of observing time that enabled this large project.
\Kepler\ was competitively selected as the tenth NASA Discovery mission. Funding for this mission is provided by the NASA Science Mission Directorate.  
LMW acknowledges support from the Parrent Fellowship, the Trottier Family, and the Levy Family.
AWH acknowledges NASA grant NNX12AJ23G.  
EAP acknowledges support from Hubble Fellowship grant HST-HF2-51365.001-A awarded by the Space Telescope Science Institute, which is operated by the Association of Universities for Research in Astronomy, Inc. for NASA under contract NAS 5-26555. 
LH acknowledges National Science Foundation grant AST-1009810.
ES is supported by a post-graduate scholarship from the Natural Sciences and Engineering Research Council of Canada.
PAC acknowledges National Science Foundation grant AST-1109612.
Finally, the authors wish to recognize and acknowledge the very significant cultural role and reverence that the summit of Maunakea has always had within the indigenous Hawaiian community.  We are most fortunate to have the opportunity to conduct observations from this mountain.

\facilities{Keck:I (HIRES), Kepler}
\software{python2.7, scipy, pandas}

\bibliography{cks-bib}

\begin{thebibliography}{}
\expandafter\ifx\csname natexlab\endcsname\relax\def\natexlab#1{#1}\fi
\providecommand{\url}[1]{\href{#1}{#1}}
\providecommand{\dodoi}[1]{doi:~\href{http://doi.org/#1}{\nolinkurl{#1}}}
\providecommand{\doeprint}[1]{\href{http://ascl.net/#1}{\nolinkurl{http://ascl.net/#1}}}
\providecommand{\doarXiv}[1]{\href{https://arxiv.org/abs/#1}{\nolinkurl{https://arxiv.org/abs/#1}}}

\bibitem[{{Ballard} \& {Johnson}(2016)}]{Ballard2016}
{Ballard}, S., \& {Johnson}, J.~A. 2016, \apj, 816, 66,
  \dodoi{10.3847/0004-637X/816/2/66}

\bibitem[{{Becker} {et~al.}(2015){Becker}, {Vanderburg}, {Adams}, {Rappaport},
  \& {Schwengeler}}]{Becker2015}
{Becker}, J.~C., {Vanderburg}, A., {Adams}, F.~C., {Rappaport}, S.~A., \&
  {Schwengeler}, H.~M. 2015, \apjl, 812, L18,
  \dodoi{10.1088/2041-8205/812/2/L18}

\bibitem[{Borucki {et~al.}(2010)Borucki, Koch, Brown, Basri, Batalha, Caldwell,
  Cochran, Dunham, {Gautier III}, Geary, Gilliland, Howell, Jenkins, Latham,
  Lissauer, Marcy, Monet, Rowe, \& Sasselov}]{Borucki2010}
Borucki, W.~J., Koch, D.~G., Brown, T.~M., {et~al.} 2010, \apjl, 713, L126,
  \dodoi{10.1088/2041-8205/713/2/L126}

\bibitem[{Brown {et~al.}(2011)Brown, Latham, Everett, \& Esquerdo}]{Brown2011}
Brown, T.~M., Latham, D.~W., Everett, M.~E., \& Esquerdo, G.~a. 2011, The
  Astronomical Journal, 142, 112, \dodoi{10.1088/0004-6256/142/4/112}

\bibitem[{{Butler} {et~al.}(1999){Butler}, {Marcy}, {Fischer}, {Brown},
  {Contos}, {Korzennik}, {Nisenson}, \& {Noyes}}]{Butler1999}
{Butler}, R.~P., {Marcy}, G.~W., {Fischer}, D.~A., {et~al.} 1999, \apj, 526,
  916, \dodoi{10.1086/308035}

\bibitem[{{Chatterjee} {et~al.}(2008){Chatterjee}, {Ford}, {Matsumura}, \&
  {Rasio}}]{Chatterjee2008}
{Chatterjee}, S., {Ford}, E.~B., {Matsumura}, S., \& {Rasio}, F.~A. 2008, \apj,
  686, 580, \dodoi{10.1086/590227}

\bibitem[{{Dai} {et~al.}(2018){Dai}, {Masuda}, \& {Winn}}]{Dai2018}
{Dai}, F., {Masuda}, K., \& {Winn}, J.~N. 2018, ArXiv e-prints,
  arXiv:1808.08475.
\newblock \doarXiv{1808.08475}

\bibitem[{{Dawson} {et~al.}(2016){Dawson}, {Lee}, \& {Chiang}}]{Dawson2016}
{Dawson}, R.~I., {Lee}, E.~J., \& {Chiang}, E. 2016, \apj, 822, 54,
  \dodoi{10.3847/0004-637X/822/1/54}

\bibitem[{{Fabrycky} {et~al.}(2014){Fabrycky}, {Lissauer}, {Ragozzine}, {Rowe},
  {Steffen}, {Agol}, {Barclay}, {Batalha}, {Borucki}, {Ciardi}, {Ford},
  {Gautier}, {Geary}, {Holman}, {Jenkins}, {Li}, {Morehead}, {Morris},
  {Shporer}, {Smith}, {Still}, \& {Van Cleve}}]{Fabrycky2014}
{Fabrycky}, D.~C., {Lissauer}, J.~J., {Ragozzine}, D., {et~al.} 2014, \apj,
  790, 146, \dodoi{10.1088/0004-637X/790/2/146}

\bibitem[{Fang \& Margot(2012)}]{Fang2012}
Fang, J., \& Margot, J.-L. 2012, \apj, 761, 92,
  \dodoi{10.1088/0004-637X/761/2/92}

\bibitem[{{Fischer} \& {Valenti}(2005)}]{Fischer2005}
{Fischer}, D.~A., \& {Valenti}, J. 2005, \apj, 622, 1102,
  \dodoi{10.1086/428383}

\bibitem[{{Ford} {et~al.}(2011){Ford}, {Rowe}, {Fabrycky}, {Carter}, {Holman},
  {Lissauer}, {Ragozzine}, {Steffen}, {Batalha}, {Borucki}, {Bryson},
  {Caldwell}, {Dunham}, {Gautier}, {Jenkins}, {Koch}, {Li}, {Lucas}, {Marcy},
  {McCauliff}, {Mullally}, {Quintana}, {Still}, {Tenenbaum}, {Thompson}, \&
  {Twicken}}]{Ford2011}
{Ford}, E.~B., {Rowe}, J.~F., {Fabrycky}, D.~C., {et~al.} 2011, \apjs, 197, 2,
  \dodoi{10.1088/0067-0049/197/1/2}

\bibitem[{Fressin {et~al.}(2013)Fressin, Torres, Charbonneau, Bryson,
  Christiansen, Dressing, Jenkins, Walkowicz, \& Batalha}]{Fressin2013}
Fressin, F., Torres, G., Charbonneau, D., {et~al.} 2013, The Astrophysical
  Journal, 766, 81

\bibitem[{{Fulton} {et~al.}(2017){Fulton}, {Petigura}, {Howard}, {Isaacson},
  {Marcy}, {Cargile}, {Hebb}, {Weiss}, {Johnson}, {Morton}, {Sinukoff},
  {Crossfield}, \& {Hirsch}}]{Fulton2017}
{Fulton}, B.~J., {Petigura}, E.~A., {Howard}, A.~W., {et~al.} 2017, \aj, 154,
  109, \dodoi{10.3847/1538-3881/aa80eb}

\bibitem[{{Furlan} {et~al.}(2017){Furlan}, {Ciardi}, {Everett}, {Saylors},
  {Teske}, {Horch}, {Howell}, {van Belle}, {Hirsch}, {Gautier}, {Adams},
  {Barrado}, {Cartier}, {Dressing}, {Dupree}, {Gilliland}, {Lillo-Box},
  {Lucas}, \& {Wang}}]{Furlan2017}
{Furlan}, E., {Ciardi}, D.~R., {Everett}, M.~E., {et~al.} 2017, \aj, 153, 71,
  \dodoi{10.3847/1538-3881/153/2/71}

\bibitem[{{Gaia Collaboration} {et~al.}(2018){Gaia Collaboration}, {Brown},
  {Vallenari}, {Prusti}, {de Bruijne}, {Babusiaux}, \&
  {Bailer-Jones}}]{GaiaDR2}
{Gaia Collaboration}, {Brown}, A.~G.~A., {Vallenari}, A., {et~al.} 2018, ArXiv
  e-prints.
\newblock \doarXiv{1804.09365}

\bibitem[{{Gaidos} {et~al.}(2016){Gaidos}, {Mann}, {Kraus}, \&
  {Ireland}}]{Gaidos2016}
{Gaidos}, E., {Mann}, A.~W., {Kraus}, A.~L., \& {Ireland}, M. 2016, \mnras,
  457, 2877, \dodoi{10.1093/mnras/stw097}

\bibitem[{{Ginzburg} {et~al.}(2018){Ginzburg}, {Schlichting}, \&
  {Sari}}]{Ginzburg2018}
{Ginzburg}, S., {Schlichting}, H.~E., \& {Sari}, R. 2018, \mnras, 476, 759,
  \dodoi{10.1093/mnras/sty290}

\bibitem[{Hansen \& Murray(2013)}]{Hansen2013}
Hansen, B. M.~S., \& Murray, N. 2013, The Astrophysical Journal, 775, 53,
  \dodoi{10.1088/0004-637X/775/1/53}

\bibitem[{Johnson {et~al.}(2007)Johnson, Butler, Marcy, Fischer, Vogt, Wright,
  \& Peek}]{Johnson2007}
Johnson, J.~A., Butler, R.~P., Marcy, G.~W., {et~al.} 2007, The Astrophysical
  Journal, 670, 833

\bibitem[{{Johnson} {et~al.}(2017){Johnson}, {Petigura}, {Fulton}, {Marcy},
  {Howard}, {Isaacson}, {Hebb}, {Cargile}, {Morton}, {Weiss}, {Winn}, {Rogers},
  {Sinukoff}, \& {Hirsch}}]{Johnson2017}
{Johnson}, J.~A., {Petigura}, E.~A., {Fulton}, B.~J., {et~al.} 2017, \aj, 154,
  108, \dodoi{10.3847/1538-3881/aa80e7}

\bibitem[{{Latham} {et~al.}(2011){Latham}, {Rowe}, {Quinn}, {Batalha},
  {Borucki}, {Brown}, {Bryson}, {Buchhave}, {Caldwell}, {Carter},
  {Christiansen}, {Ciardi}, {Cochran}, {Dunham}, {Fabrycky}, {Ford}, {Gautier},
  {Gilliland}, {Holman}, {Howell}, {Ibrahim}, {Isaacson}, {Jenkins}, {Koch},
  {Lissauer}, {Marcy}, {Quintana}, {Ragozzine}, {Sasselov}, {Shporer},
  {Steffen}, {Welsh}, \& {Wohler}}]{Latham2011}
{Latham}, D.~W., {Rowe}, J.~F., {Quinn}, S.~N., {et~al.} 2011, \apjl, 732, L24,
  \dodoi{10.1088/2041-8205/732/2/L24}

\bibitem[{{Lee} \& {Chiang}(2017)}]{Lee2017}
{Lee}, E.~J., \& {Chiang}, E. 2017, \apj, 842, 40,
  \dodoi{10.3847/1538-4357/aa6fb3}

\bibitem[{{Lissauer} {et~al.}(2011){Lissauer}, {Ragozzine}, {Fabrycky},
  {Steffen}, {Ford}, {Jenkins}, {Shporer}, {Holman}, {Rowe}, {Quintana},
  {Batalha}, {Borucki}, {Bryson}, {Caldwell}, {Carter}, {Ciardi}, {Dunham},
  {Fortney}, {Gautier}, {Howell}, {Koch}, {Latham}, {Marcy}, {Morehead}, \&
  {Sasselov}}]{Lissauer2011_multis}
{Lissauer}, J.~J., {Ragozzine}, D., {Fabrycky}, D.~C., {et~al.} 2011, \apjs,
  197, 8, \dodoi{10.1088/0067-0049/197/1/8}

\bibitem[{Lissauer {et~al.}(2012)Lissauer, Marcy, Rowe, Bryson, Adams,
  Buchhave, Ciardi, Cochran, Fabrycky, Ford, Fressin, Geary, Gilliland, Holman,
  Howell, Jenkins, Kinemuchi, Koch, Morehead, Ragozzine, Seader, Tanenbaum,
  Torres, \& Twicken}]{Lissauer2012}
Lissauer, J.~J., Marcy, G.~W., Rowe, J.~F., {et~al.} 2012, \apj, 750, 112,
  \dodoi{10.1088/0004-637X/750/2/112}

\bibitem[{Lissauer {et~al.}(2014)Lissauer, Marcy, Bryson, Rowe, Jontof-Hutter,
  Agol, Borucki, Carter, Ford, Gilliland, Kolbl, Star, Steffen, \&
  Torres}]{Lissauer2014}
Lissauer, J.~J., Marcy, G.~W., Bryson, S.~T., {et~al.} 2014, \apj, 784, 44,
  \dodoi{10.1088/0004-637X/784/1/44}

\bibitem[{{Mulders} {et~al.}(2018){Mulders}, {Pascucci}, {Apai}, \&
  {Ciesla}}]{Mulders2018}
{Mulders}, G.~D., {Pascucci}, I., {Apai}, D., \& {Ciesla}, F.~J. 2018, \aj,
  156, 24, \dodoi{10.3847/1538-3881/aac5ea}

\bibitem[{Mullally {et~al.}(2015)Mullally, Coughlin, Thompson, Rowe, Burke,
  Latham, Batalha, Bryson, Christiansen, Henze, Ofir, Quarles, Shporer, Eylen,
  Laerhoven, Shah, Wolfgang, Chaplin, Xie, Akeson, Argabright, Bachtell,
  Barclay, Borucki, Caldwell, Campbell, Catanzarite, Cochran, Duren, Fleming,
  Fraquelli, Girouard, Haas, Hełminiak, Howell, Huber, Larson, III, Jenkins,
  Li, Lissauer, McArthur, Miller, Morris, Patil-Sabale, Plavchan, Putnam,
  Quintana, Ramirez, Aguirre, Seader, Smith, Steffen, Stewart, Stober, Still,
  Tenenbaum, Troeltzsch, Twicken, \& Zamudio}]{Mullally2015}
Mullally, F., Coughlin, J.~L., Thompson, S.~E., {et~al.} 2015, The
  Astrophysical Journal Supplement Series, 217, 31,
  \dodoi{10.1088/0067-0049/217/2/31}

\bibitem[{{Munoz Romero} \& {Kempton}(2018)}]{MunozRomero2018}
{Munoz Romero}, C.~E., \& {Kempton}, E.~M.-R. 2018, \aj, 155, 134,
  \dodoi{10.3847/1538-3881/aaab5e}

\bibitem[{{Obertas} {et~al.}(2017){Obertas}, {Van Laerhoven}, \&
  {Tamayo}}]{Obertas2017}
{Obertas}, A., {Van Laerhoven}, C., \& {Tamayo}, D. 2017, \icarus, 293, 52,
  \dodoi{10.1016/j.icarus.2017.04.010}

\bibitem[{{Owen} \& {Wu}(2017)}]{Owen2017}
{Owen}, J.~E., \& {Wu}, Y. 2017, ArXiv e-prints.
\newblock \doarXiv{1705.10810}

\bibitem[{{Petigura}(2015)}]{Petigura2015PhD}
{Petigura}, E.~A. 2015, PhD thesis, University of California, Berkeley

\bibitem[{Petigura {et~al.}(2013)Petigura, Howard, \& Marcy}]{Petigura2013}
Petigura, E.~A., Howard, A.~W., \& Marcy, G.~W. 2013, Proceedings of the
  National Academy of Sciences, 110, 19273, \dodoi{10.1073/pnas.1319909110}

\bibitem[{{Petigura} {et~al.}(2017){Petigura}, {Howard}, {Marcy}, {Johnson},
  {Isaacson}, {Cargile}, {Hebb}, {Fulton}, {Weiss}, {Morton}, {Winn}, {Rogers},
  {Sinukoff}, {Hirsch}, \& {Crossfield}}]{Petigura2017}
{Petigura}, E.~A., {Howard}, A.~W., {Marcy}, G.~W., {et~al.} 2017, \aj, 154,
  107, \dodoi{10.3847/1538-3881/aa80de}

\bibitem[{{Petigura} {et~al.}(2018){Petigura}, {Marcy}, {Winn}, {Weiss},
  {Fulton}, {Howard}, {Sinukoff}, {Isaacson}, {Morton}, \&
  {Johnson}}]{Petigura2018}
{Petigura}, E.~A., {Marcy}, G.~W., {Winn}, J.~N., {et~al.} 2018, \aj, 155, 89,
  \dodoi{10.3847/1538-3881/aaa54c}

\bibitem[{{Petrovich} {et~al.}(2018){Petrovich}, {Deibert}, \&
  {Wu}}]{Petrovich2018}
{Petrovich}, C., {Deibert}, E., \& {Wu}, Y. 2018, ArXiv e-prints,
  arXiv:1804.05065.
\newblock \doarXiv{1804.05065}

\bibitem[{Rogers(2015)}]{Rogers2015}
Rogers, L.~a. 2015, \apj, 801, 41, \dodoi{10.1088/0004-637X/801/1/41}

\bibitem[{{Rowe} {et~al.}(2014){Rowe}, {Bryson}, {Marcy}, {Lissauer},
  {Jontof-Hutter}, {Mullally}, {Gilliland}, {Issacson}, {Ford}, {Howell},
  {Borucki}, {Haas}, {Huber}, {Steffen}, {Thompson}, {Quintana}, {Barclay},
  {Still}, {Fortney}, {Gautier}, {Hunter}, {Caldwell}, {Ciardi}, {Devore},
  {Cochran}, {Jenkins}, {Agol}, {Carter}, \& {Geary}}]{Rowe2014}
{Rowe}, J.~F., {Bryson}, S.~T., {Marcy}, G.~W., {et~al.} 2014, \apj, 784, 45,
  \dodoi{10.1088/0004-637X/784/1/45}

\bibitem[{{Sanchis-Ojeda} {et~al.}(2014){Sanchis-Ojeda}, {Rappaport}, {Winn},
  {Kotson}, {Levine}, \& {El Mellah}}]{Sanchis-Ojeda2014}
{Sanchis-Ojeda}, R., {Rappaport}, S., {Winn}, J.~N., {et~al.} 2014, \apj, 787,
  47, \dodoi{10.1088/0004-637X/787/1/47}

\bibitem[{{Santerne} {et~al.}(2016){Santerne}, {Moutou}, {Tsantaki}, {Bouchy},
  {H{\'e}brard}, {Adibekyan}, {Almenara}, {Amard}, {Barros}, {Boisse},
  {Bonomo}, {Bruno}, {Courcol}, {Deleuil}, {Demangeon}, {D{\'{\i}}az},
  {Guillot}, {Havel}, {Montagnier}, {Rajpurohit}, {Rey}, \&
  {Santos}}]{Santerne2016}
{Santerne}, A., {Moutou}, C., {Tsantaki}, M., {et~al.} 2016, \aap, 587, A64,
  \dodoi{10.1051/0004-6361/201527329}

\bibitem[{{Spalding} \& {Batygin}(2016)}]{Spalding2016}
{Spalding}, C., \& {Batygin}, K. 2016, \apj, 830, 5,
  \dodoi{10.3847/0004-637X/830/1/5}

\bibitem[{{Steffen} \& {Coughlin}(2016)}]{Steffen2016}
{Steffen}, J.~H., \& {Coughlin}, J.~L. 2016, Proceedings of the National
  Academy of Science, 113, 12023, \dodoi{10.1073/pnas.1606658113}

\bibitem[{{Steffen} {et~al.}(2012){Steffen}, {Ragozzine}, {Fabrycky}, {Carter},
  {Ford}, {Holman}, {Rowe}, {Welsh}, {Borucki}, {Boss}, {Ciardi}, \&
  {Quinn}}]{Steffen2012}
{Steffen}, J.~H., {Ragozzine}, D., {Fabrycky}, D.~C., {et~al.} 2012,
  Proceedings of the National Academy of Science, 109, 7982,
  \dodoi{10.1073/pnas.1120970109}

\bibitem[{{Tamayo} {et~al.}(2018){Tamayo}, {Hadden}, {Hussain}, {Silburt},
  {Gilbertson}, {Rein}, \& {Menou}}]{Tamayo2018}
{Tamayo}, D., {Hadden}, S., {Hussain}, N., {et~al.} 2018, in AAS/Division of
  Dynamical Astronomy Meeting, 201.02

\bibitem[{{Tremaine} \& {Dong}(2012)}]{Tremaine2012}
{Tremaine}, S., \& {Dong}, S. 2012, \aj, 143, 94,
  \dodoi{10.1088/0004-6256/143/4/94}

\bibitem[{{Van Eylen} {et~al.}(2018){Van Eylen}, {Albrecht}, {Huang},
  {MacDonald}, {Dawson}, {Cai}, {Foreman-Mackey}, {Lundkvist}, {Silva Aguirre},
  {Snellen}, \& {Winn}}]{vanEylen2018_ecc}
{Van Eylen}, V., {Albrecht}, S., {Huang}, X., {et~al.} 2018, ArXiv e-prints,
  arXiv:1807.00549.
\newblock \doarXiv{1807.00549}

\bibitem[{Weiss \& Marcy(2014)}]{Weiss2014}
Weiss, L.~M., \& Marcy, G.~W. 2014, \apj, 783, L6,
  \dodoi{10.1088/2041-8205/783/1/L6}

\bibitem[{{Weiss} {et~al.}(2018){Weiss}, {Marcy}, {Petigura}, {Fulton},
  {Howard}, {Winn}, {Isaacson}, {Morton}, {Hirsch}, {Sinukoff}, {Cumming},
  {Hebb}, \& {Cargile}}]{Weiss2018}
{Weiss}, L.~M., {Marcy}, G.~W., {Petigura}, E.~A., {et~al.} 2018, \aj, 155, 48,
  \dodoi{10.3847/1538-3881/aa9ff6}

\bibitem[{Wright {et~al.}(2009)Wright, Upadhyay, Marcy, Fischer, Ford, \&
  Johnson}]{Wright2009}
Wright, J.~T., Upadhyay, S., Marcy, G.~W., {et~al.} 2009, The Astrophysical
  Journal, 693, 1084, \dodoi{10.1088/0004-637X/693/2/1084}

\bibitem[{{Xie} {et~al.}(2016){Xie}, {Dong}, {Zhu}, {Huber}, {Zheng}, {De Cat},
  {Fu}, {Liu}, {Luo}, {Wu}, {Zhang}, {Zhang}, {Zhou}, {Cao}, {Hou}, {Wang}, \&
  {Zhang}}]{Xie2016}
{Xie}, J.-W., {Dong}, S., {Zhu}, Z., {et~al.} 2016, Proceedings of the National
  Academy of Science, 113, 11431, \dodoi{10.1073/pnas.1604692113}

\bibitem[{{Zhu} {et~al.}(2018){Zhu}, {Petrovich}, {Wu}, {Dong}, \&
  {Xie}}]{Zhu2018}
{Zhu}, W., {Petrovich}, C., {Wu}, Y., {Dong}, S., \& {Xie}, J. 2018, ArXiv
  e-prints.
\newblock \doarXiv{1802.09526}

\end{thebibliography}
\end{document}